\documentclass[aps,prd,twocolumn,showpacs,preprintnumbers,nofootinbib,amsmath,amssymb]{revtex4}
\usepackage{graphicx}
\usepackage{dcolumn}
\usepackage{bm}
\usepackage{amsmath}
\usepackage{braket}
\usepackage[normalem]{ulem}
\usepackage[hidelinks]{hyperref}
\usepackage{epstopdf}
\usepackage{placeins}
\usepackage[normalem]{ulem}

\newcommand{\beq}{\begin{eqnarray}}
\newcommand{\eeq}{\end{eqnarray}}

\newcommand{\Tr}{\ensuremath{\mathrm{Tr}}}

\def\spose#1{\hbox to 0pt{#1\hss}}
\def\ltapprox{\mathrel{\spose{\lower 3pt\hbox{$\mathchar"218$}}
 \raise 2.0pt\hbox{$\mathchar"13C$}}}

\begin{document}

\title{Topological properties of $CP^{N-1}$ models in the large-$N$ limit}

\author{Claudio Bonanno}
\email{claudio.bonanno@pi.infn.it}
\affiliation{Universit\`a di Pisa and INFN Sezione di Pisa,\\ 
Largo Pontecorvo 3, I-56127 Pisa, Italy
}
\author{Claudio Bonati}
\email{claudio.bonati@df.unipi.it}
\affiliation{Universit\`a di Pisa and INFN Sezione di Pisa,\\ 
Largo Pontecorvo 3, I-56127 Pisa, Italy
}
\author{Massimo D'Elia}
\email{massimo.delia@unipi.it}
\affiliation{Universit\`a di Pisa and INFN Sezione di Pisa,\\ 
Largo Pontecorvo 3, I-56127 Pisa, Italy
}

\date{\today}

\begin{abstract}
We investigate, by numerical simulations on a lattice, the $\theta$-dependence
of 2$d$ $CP^{N-1}$ models for a range of $N$ going from 9 to 31, combining
imaginary $\theta$ and simulated tempering techniques to improve the
signal-to-noise ratio and alleviate the critical slowing down of the
topological modes.  We provide continuum extrapolations for the second and
fourth order coefficients in the Taylor expansion in $\theta$ of the vacuum
energy of the theory, parameterized in terms of the topological susceptibility
$\chi$ and of the so-called $b_2$ coefficient. Those are then compared with
available analytic predictions obtained within the $1/N$ expansion, pointing
out that higher order corrections might be relevant in the explored range of
$N$, and that this fact might be related to the non-analytic behavior expected
for $N = 2$.  We also consider sixth-order corrections in the $\theta$
expansion, parameterized in terms of the so-called $b_4$ coefficient: in this
case  our present statistical accuracy permits to have reliable non-zero
continuum estimations only for $N \leq 11$, while for  larger values we can
only set upper bounds.  The sign and values obtained for $b_4$ are compared to
large-$N$ predictions, as well as to results obtained for $SU(N_c)$ Yang-Mills
theories, for which a first numerical determination is provided in this study
for the case $N_c = 2$.
\end{abstract}

\maketitle

\section{Introduction}
\label{section_intro}

The 2$d$ $CP^{N-1}$ models are a class of quantum field theories which have
gained much importance in the study of non-perturbative properties of quantum
gauge theories. Indeed, they present many interesting non-perturbative
features, analogous to the ones of Yang-Mills theories, such as the existence
of a $\theta$-term, the presence of topologically-stable instantonic
configurations and confinement of fundamental matter fields~\cite{witten,
cpn_article_confinement,cpn_general_properties,advanced_topics_QFT}.  These
characteristics make the $CP^{N-1}$ theories ideal test-beds for the study of
the main non-perturbative properties of gauge theories.

The Euclidean action of the $CP^{N-1}$ models, with the topological term,
can be written, introducing a non-propagating abelian gauge field $A_\mu$, as
\begin{equation}
S(\theta)= \int \bigg[\frac{N}{g} \bar{D}_\mu \bar{z}(x) D_\mu z(x) -i \theta q(x) \bigg]d^2x,
\end{equation}
where $z$ is a complex $N$-component scalar field satisfying
$\bar{z}(x)z(x)=1$, $D_\mu$ is the usual $U(1)$ covariant derivative, $g$ is
the 't Hooft coupling, $\theta$ is the $CP$-violating angle and
\begin{equation}
Q=\int q(x) d^2x =\frac{1}{4\pi}\epsilon_{\mu\nu}\int F_{\mu\nu}(x) d^2x
\end{equation}
is the topological charge.

Like $SU(N_c)$ models, also $CP^{N-1}$ models offer the possibility of performing a
large-$N$ 't Hooft limit, i.e. keeping $g$ fixed, to study the non-perturbative
regime. However, contrary to Yang-Mills theories, these models are exactly
solvable in this limit and large-$N$ predictions are also 
quantitative~\cite{calcolo_e_2,SU(N)_large_N_limit,cpn_large_N,SU(N)_theta_dep_manca,cpn_next_to_leading_no_susc}.
Therefore, a careful inspection of this limit is possible and deserved in order
to better understand the non-perturbative behavior of quantum gauge theories.

One of the most intriguing non-perturbative properties of $CP^{N-1}$ models is
the $\theta$-dependence of their vacuum energy (density), defined as:
\begin{equation}
E(\theta) \equiv -\frac{\log Z(\theta)}{V} = -\frac{1}{V}\log \int [dA][d\bar{z}][dz] e^{-S(\theta)}
\end{equation}
where $V$ is the 2$d$ space-time volume. The vacuum energy is an even function
of $\theta$ and is assumed to be analytic around $\theta=0$, so that one can
express it as a Taylor expansion around this value.  The usual parametrization
employed in the literature is:
\begin{equation}
f(\theta) \equiv E(\theta) - E(0) = \frac{1}{2}\chi \theta^2\bigg(1+\sum_{n=0}^{\infty}b_{2n}\theta^{2n}\bigg),
\label{taylexp}
\end{equation}
where $\chi$ is the topological susceptibility and the $b_{2n}$ coefficients,
which parameterize the non-quadratic dependence of $f$ on $\theta$, are related
to ratios involving higher cumulants of the topological charge distribution
$P(Q)$.  The explicit expression of the first few terms as a function of the
cumulants $k_n$ of $P(Q)$ reads:
\begin{gather}
\chi= \frac{1}{V} k_2\bigg\vert_{\theta=0}=\frac{1}{V}\braket{Q^2}\bigg\vert_{\theta=0}\nonumber \\
b_2=-\frac{k_4}{12 \text{ } k_2}\bigg\vert_{\theta=0}
=\frac{-\braket{Q^4}+3\braket{Q^2}^2}{12 \braket{Q^2}}\bigg\vert_{\theta=0} \label{definition_chi_b_2n} \\
b_4=\frac{k_6}{360 \text{ } k_2}\bigg\vert_{\theta=0}=
\frac{\braket{Q^6}-15\braket{Q^4}\braket{Q^2}+30\braket{Q^2}^3}{360 \braket{Q^2}}\bigg\vert_{\theta=0} \nonumber
\end{gather}

On general grounds, large-$N$ arguments predict
$\chi=\bar{\chi}N^{-1}+O(N^{-2})$ and
$b_{2n}=\bar{b}_{2n}N^{-2n}+O(N^{-2n-1})$. More precisely, one can show that
\cite{calcolo_e_2, SU(N)_theta_dep_manca, SU(N)_large_N_limit, cpn_large_N}
\begin{gather} \label{large_N}
\xi^2 \chi=\frac{1}{2\pi N}+\frac{e_2}{N^2}+O\bigg(\frac{1}{N^3}\bigg),\\
b_2=-\frac{27}{5}\frac{1}{N^2}+O\bigg(\frac{1}{N^3}\bigg),
\label{b2largen}\\
b_4=-\frac{25338}{175}\frac{1}{N^4}+O\bigg(\frac{1}{N^5}\bigg),
\end{gather}
that $\bar{b}_{2n}<0$ and that $e_2 \simeq
-0.0606$~\cite{calcolo_e_2}.  The length scale $\xi$ appearing in
Eq.~(\ref{large_N}) is the second moment correlation length, defined as:
\begin{equation}\label{def_xi}
\xi^2 \equiv \frac{1}{\int G(x)d^2x}\int G(x) \frac{\vert x\vert^2}{4} d^2 x,
\end{equation}
where
\begin{equation}\label{projector_definition}
G(x) \equiv \braket{P_{ij}(x)P_{ij}(0)}-\frac{1}{N}, \quad P_{ij}(x) \equiv z_i(x) \bar{z}_j(x).
\end{equation}
If on one hand many efforts were put in the analytic study of
the large-$N$ limit of the $CP^{N-1}$ models, on the other hand present
numerical results are quite limited for these theories and mostly regarding the
topological susceptibility.  The situation is in some sense opposite to the case of
$SU(N_c)$ pure gauge theories, where quantitative analytic large-$N$
predictions are lacking, but numerical results for the large-$N$ behavior are
available both for $\chi$~\cite{SU(N)_theta_dep_vicari, Lucini:2004yh,
Ce:2016awn, SU(N)_large_N_limit} and $b_2$~\cite{SU(N)_theta_dep_vicari,
DElia:2003zne, Giusti:2007tu, Bonati:2013tt, Ce:2015qha, theta_dep_SU3_d_elia,
SU(N)_large_N_limit}.
This fact constitutes a strong motivation to make a similar numerical 
inspection for the $CP^{N-1}$ theories too and to look for a numerical 
validation of large-$N$ analytic predictions for these models.

The leading term in the $1/N$ expansion of the topological
susceptibility has been checked quite carefully on the lattice
\cite{MC_simulation_cpn, fighting_slowing_down_cpn, susc_top_cpn,
renormalization_lattice_top_charge_1}, however large uncertainties exist
regarding the next-to-leading correction.  Indeed, many numerical works on the
$CP^{N-1}$ theories show a deviation from the leading term which appears to be
of opposite sign compared to that predicted by Eq.~(\ref{large_N})
\cite{fighting_slowing_down_cpn, vicari_review_QCD_CPN}. As for the
$O(\theta^4)$ term only a preliminary investigation is
available~\cite{tesi_manca}, while for the other high-order corrections
no numerical results
have been reported up to now.

The purpose of the present paper is to make progress in this direction. For
this reason we consider a range of $N$ going from 9 to 31, with the aim of
determining the first coefficients in the $1/N$ expansion for at least $\chi$
and $b_2$ and, possibly, for $b_4$ too. Two
main numerical difficulties have to be faced to complete this task:

{\em i)} as the continuum limit and/or the large-$N$ limit are approached,
standard updating algorithms fail to correctly sample the distribution of $Q$
and field configurations get trapped in path integral sectors with fixed
topology. This problem is usually known as the {\em freezing} of topological
modes~\cite{Alles:1996vn, critical_slowing_down_review, Schaefer:2010hu,
Bonati:2017woi};

{\em ii)} the determination of higher order cumulants usually requires the
detection of slight deviations of $P(Q)$ from a Gaussian distribution,
resulting in a low statistical accuracy (for a more detailed discussion of this problem see, e.g.,
Ref.~\cite{theta_dep_SU3_d_elia}). The problem is more and more critical as the
large-$N$ limit is approached, given that $b_{2n}$ is expected to vanish like
$1/N^{2n}$.

To alleviate the second problem, we consider analytic continuation from
simulations performed at imaginary $\theta$ values.  Analytic continuation,
originally explored for the study of QCD at finite baryon chemical
potential~\cite{deForcrand:2002hgr, DElia:2002tig}, has been introduced as a
tool to explore the $\theta$-dependence of the mass gap in $CP^1$
models~\cite{Alles:2007br,Alles:2008wm,Alles:2014tta} and then successfully
extended to the investigation of $SU(N_c)$ gauge theories at non-zero
$\theta$~\cite{Aoki:2008gv, firts_article_theta_immaginario_SUN, DElia:2012pvq,
DElia:2013uaf, theta_dep_SU3_d_elia, Guo:2015tla, SU(N)_large_N_limit,
Bonati:2018rfg}, there leading to a substantial improvement in the
determination of the higher-order coefficients of $f(\theta)$, compared to the
measure at $\theta=0$ alone.
 
As for the first problem, we have decided to exploit simulated
tempering~\cite{vicari_simul_temp_cpn, simul_temp_first_article}: although
other methods have been proposed recently
\cite{fighting_slowing_down_cpn, meta_surf_top_barriers, worm_algorithm}, this
algorithm has proven to be sufficiently powerful to allow us to reasonably
contain statistical errors in the explored range of $N$.

The paper is organized as follows. In Sec.~\ref{section_setup} we
present the adopted discretization for the action and for the topological
charge, along with a description of the imaginary-$\theta$ method and of the
simulated tempering algorithm. In Sec.~\ref{section_results} we present our
numerical results for $\chi$, $b_2$ and $b_4$.  Results are then discussed and
compared to large-$N$ analytic predictions in Sec.~\ref{results_chi}: there we
discuss about the possible influence from higher order terms in the $1/N$
expansion, also in connection with the non-analytic behavior expected for $N =
2$; moreover, results obtained for $b_4$ are compared also with a determination
of this coefficient in Yang-Mills theories that we provide for the first time
in this study. Finally, in Sec.~\ref{section_conclusions}, we draw our
conclusions.

\section{Numerical Set-up}
\label{section_setup}

\subsection{Lattice action}\label{subsection_lattice_action}

We discretized the theory on a square lattice of size
$L$ with periodic boundary conditions. For the non-topological part of the
action, we adopted the tree-level $O(a)$ Symanzik-improved lattice action
\cite{MC_simulation_cpn}
\begin{equation}\label{Symanzik_improved_lattice_action_cpn}
\begin{gathered}
S_L=-2N\beta_L\sum_{x,\mu}\bigg\{ c_1\Re\bigg[\bar{U}_\mu(x)\bar{z}(x+\hat{\mu})z(x)\bigg]-c_1 \\
+c_2\Re\bigg[\bar{U}_\mu(x+\hat{\mu})\bar{U}_\mu(x)\bar{z}(x+2\hat{\mu})z(x)\bigg]-c_2\bigg\},
\end{gathered}
\end{equation}
where $c_1=4/3$, $c_2=-1/12$, $\beta_L \equiv 1/g_L$ is the bare coupling and
$U_\mu(x)$ satisfies $\bar{U}_\mu(x)U_\mu(x)=1$. 
Next-to-nearest-neighbor interactions are introduced to cancel $O(a)$
corrections to the continuum limit.

Several discretizations of the topological charge exist: the most
straightforward one makes use of the plaquette operator
$\Pi_{\mu\nu}(x)$:
\begin{equation}\label{def_non_geo_charge}
\begin{aligned}
Q_L=\frac{1}{4\pi}\sum_{x,\mu,\nu} \epsilon_{\mu\nu}\Im\bigg[\Pi_{\mu\nu}(x)\bigg]\\
=\frac{1}{2\pi}\sum_x \Im\bigg[\Pi_{12}(x)\bigg],
\end{aligned}
\end{equation}
where, as usual,
\begin{equation}
\Pi_{\mu \nu} (x) \equiv U_\mu(x) U_\nu(x+\hat{\mu})\bar{U}_\mu(x+\hat{\nu})\bar{U}_\nu(x).
\end{equation}
This choice leads to an analytic function of the gauge field but does not
result in an integer value for a generic configuration. There are, instead,
other possible definitions, known as geometric, which always result in an
integer value when assuming periodic boundary conditions. 
For the $CP^{N-1}$ models one can introduce two different geometric charges:
one~\cite{MC_simulation_cpn} makes use of the link variables $U_\mu(x)$
\begin{equation}
Q_{U} = \frac{1}{2\pi}\sum_{x} \Im \left\{ \log \bigg[\Pi_{12}(x)\bigg] \right\},
\end{equation}
the other~\cite{original_article_lattice_top_charge_triangle} 
uses the projector $P$ defined in Eq.~(\ref{projector_definition})
\begin{gather}
Q_{z} =\frac{1}{2\pi}\sum_{x} \Im \bigg\{ \log \Tr \bigg[P(x+\hat{1}+\hat{2})P(x+\hat{1})P(x)\bigg] \nonumber \\
+ \log\Tr \bigg[P(x+\hat{2})P(x+\hat{1}+\hat{2})P(x)\bigg]    \bigg\}.
\end{gather}
Since the three definitions agree already after a small amount of cooling (see
subsection~\ref{subsection_cooling}), one can use them equivalently. We adopted
the non-geometric one to discretize the $\theta$ term in the action
(see the next section) since it allows
to make use of an over-heat-bath update algorithm, while we used the
geometric $Q_U$ charge for measures, to avoid dealing with lattice
renormalizations (see subsection \ref{subsec_imm_theta} for further details).

Lastly, for the lattice correlation length, we adopted the usual
discretization~\cite{original_article_corr_length}:
\begin{equation}
\xi_L^2 = \frac{1}{4\sin^2\left(k/2\right)}\bigg[ \frac{\tilde{G}_L(0)}{\tilde{G}_L(k)}-1 \bigg],
\end{equation}
where $\tilde{G}_L(p)$ is the Fourier transform of $G_L$, the lattice version
of the two-point correlator of $P$ defined in Eq.~(\ref{projector_definition}),
and $k=2\pi/L$.  Being the $CP^{N-1}$ theories asymptotically free, the
continuum limit $a\to0$ is approached when $\beta_L\to\infty$ and, in
this limit, the lattice correlation length diverges as $\xi_L \sim
1/a$. This allows to express finite lattice spacing corrections to the
continuum limit in terms of $\xi_L$. In our case linear corrections are
removed by the adoption of the 
improved action, thus 
the expectation value of a generic observable will scale
towards the continuum limit like:
\begin{equation}\label{continuum_limit_scaling_xi_L}
\braket{\mathcal{O}}_L\left(\xi_L\right) = \braket{\mathcal{O}}_{\text{\textit{cont}}} + c\text{ } \xi_L^{-2} + 
O (\xi_L^{-3}) .
\end{equation}

\subsection{Imaginary-$\theta$ method}\label{subsec_imm_theta}
\label{def_imm_theta_method}

When expressed in terms of expectation values computed at $\theta = 0$, the
coefficients of the higher order terms in the Taylor expansion, reported in
Eq.~(\ref{taylexp}), are difficult to determine because they involve small
corrections to an almost Gaussian distribution.  For this reason, the direct
insertion of a source term in the action, i.e.~a non-zero $\theta$, leads to a
significant improvement of the signal-to-noise ratio, because it permits to
determine higher order terms in Eq.~(\ref{taylexp}) from the
$\theta$-dependence of lower order cumulants, see
Refs.~\cite{firts_article_theta_immaginario_SUN, theta_dep_SU3_d_elia} for a
more detailed discussion.

However, a theory with a non-vanishing $\theta$-term cannot be
straightforwardly simulated on the lattice, due to the sign problem that arises
when considering the probability distribution of trajectories at finite
$\theta$:
\begin{equation*}
P\propto e^{i\theta Q}.
\end{equation*}
The idea is then to exploit the analytic $\theta$-dependence of the theory
around $\theta = 0$ and to continue the path-integral to imaginary $\theta$
angles:
\begin{equation*}
\theta_I \equiv i \theta \implies P \propto e^{\theta_I Q} \, .
\end{equation*}
Now the topological term is real and can be simulated on the lattice. To
discretize the $\theta$-term we chose the non-geometric lattice definition of
the topological charge in Eq.~(\ref{def_non_geo_charge}):
\begin{equation}
S_L(\theta_L) = S_L - \theta_L Q_L\, .
\end{equation}
This choice allows the use of a generalization of the over-heat-bath update
used also for the $\theta=0$ case (see subsection
\ref{subsection_numerical_algorithm}).

As anticipated above, after adding the imaginary
source term to the action we are no more measuring just the fluctuations of $Q$
but also its response to the source term: that provides a strategy to determine
$\chi$ and the $b_{2n}$ coefficients that exploits the $\theta_I$-dependence of
the cumulants {$k_n$} of the topological charge distribution. Indeed, the
explicit expression of the $\theta_I$-dependence of the first few cumulants
reads:
\begin{equation}\label{imm_theta_fit}
\begin{gathered}
\frac{k_1(\theta_I)}{V} = \chi \theta_I \big[1-2b_2 \theta_I^2+3b_4 \theta_I^4+O(\theta_I^5)\big], \\
\frac{k_2(\theta_I)}{V} = \chi \big[1-6b_2 \theta_I^2+15b_4 \theta_I^4+O(\theta_I^5)\big], \\
\frac{k_3(\theta_I)}{V} = \chi \big[-12b_2 \theta_I+60b_4 \theta_I^3+O(\theta_I^4)\big], \\
\frac{k_4(\theta_I)}{V} = \chi \big[-12b_2 +180b_4 \theta_I^2+O(\theta_I^3)\big].
\end{gathered}
\end{equation}
A global fit to Eqs.~(\ref{imm_theta_fit}) provides an improved way of
measuring $\chi$ and the $b_{2n}$ on the lattice, compared to the naive
application of Eqs.~(\ref{definition_chi_b_2n}) (see
Ref.~\cite{theta_dep_SU3_d_elia} for an explicit example). Note that, since we
choose $Q_L$ to discretize the topological term in the action, $\theta_I$ is
related to the lattice angle $\theta_L$ by a multiplicative renormalization,
which is nothing but the renormalization constant relating $Q_L$ to its
continuum counterpart~\cite{renormalization_lattice_top_charge_1}:
$\theta_I=Z_{\theta}\theta_L$. This is just a technical difficulty that can be
trivially approached by treating $Z_{\theta}$ as an additional fit parameter.
On the other hand, the choice of the cooled geometric charge $Q_U$ (see
subsection \ref{subsection_cooling}) for the measure of $k_n$ introduces no
additional renormalization.

\subsection{Numerical Algorithm}\label{subsection_numerical_algorithm}

With the chosen discretization, the total lattice action is linear in both
$U$ and $z$. More precisely, the part of the
action that depends only on a single site/link variable can be expressed as:
\begin{equation}
\tilde{S}_L[\phi] = -2N\beta_L\Re\left(\bar{\phi}F_{\phi}\right)
\end{equation}
where $\phi$ stands for $U$ or $z$ and $F_{\phi}$ is
the force. Thus, a local update algorithm can be adopted. In particular, we
chose a mixture of over-heat-bath and microcanonic over-relaxation. More
precisely, following Ref.~\cite{MC_simulation_cpn}, we performed 4 sweeps of
microcanonic for every sweep of over-heat. In the following sections we
will refer to a {\em sweep} of the whole lattice as a unit of measurement of
the Monte Carlo updating time, where sweep will refer to either microcanonic or
over-heat in the proportion reported above.

A detailed description of these two update procedures is reported in
Ref.~\cite{MC_simulation_cpn}; the only difference is that here we need to
modify the $U$-force to include the topological term:
\begin{equation*}
\begin{gathered}
{F_{U}}_\mu(x,\theta_L) = {F_{U}}_\mu(x,\theta_L=0) \\
+\epsilon_{\mu\nu} \frac{i \theta_{L}}{4\pi N\beta_L} \big[\bar{\Gamma}_{\mu}^{(+)}(x) - \bar{\Gamma}_{\mu}^{(-)}(x)  \big], \quad (\nu \ne \mu),
\end{gathered}
\end{equation*}
where $\Gamma_{\mu}^{(\pm)}(x)$ is the forward/backward staple relative to $U_\mu(x)$.

This local algorithm suffers from a severe critical slowing down of topological
modes: freezing is experienced both in the continuum limit and in the large-$N$
limit, as reported in Ref.~\cite{critical_slowing_down_review}. To overcome
this problem we exploited the simulated tempering 
algorithm~\cite{simul_temp_first_article,vicari_simul_temp_cpn}.  The main idea behind
this algorithm is to enlarge the configuration space promoting the parameters
of the probability distribution to dynamical variables. In our case, setting
$E_L\equiv S_L/\beta_L$, that means:
\begin{equation}
P[\phi] \propto e^{-\beta_L E_L[\phi]+\theta_L Q_L[\phi]} \to P[\phi,\beta_L,\theta_L]\ .
\end{equation}
Since decreasing $\beta_L$ corresponds to
increasing the lattice spacing, by changing $\beta_L$ we can move the theory away
from the continuum limit, where the slowing down is absent. This allows a
faster change in the topological charge of the configuration at the highest
$\beta_L$, which is the most affected by the freezing. As for the change of
$\theta_L$, larger angles make higher-charge configurations easier to realize.
Thus, changing $\theta_L$ constitutes a further stimulation to change the
charge of a configuration at a given $\beta_L$.

To obtain an actual benefit, it is of utmost importance that the probability of
occupying a certain couple $(\beta_L,\theta_L)$ is as uniform as
possible. To achieve this, we add a term to the action which depends
just on $\theta_L$ and $\beta_L$:
\begin{equation}
P^{\prime}[\phi,\beta_L,\theta_L] \propto e^{-\beta_L E_L[\phi]+\theta_L Q_L[\phi] + F_L(\beta_L,\theta_L)},
\end{equation}
where the optimal choice, making the distribution exactly uniform,
coincides with the free energy of the theory, $F_L(\beta_L,\theta_L)=-\log
Z_L(\beta_L,\theta_L)$. Indeed with this choice:
\begin{equation*}
\mathcal{P}^{\prime}(\beta_L,\theta_L) \propto e^{F_L(\beta_L,\theta_L)} \int [d\phi] e^{-\beta_L E_L[\phi]+\theta_L Q_L[\phi]} 
\end{equation*}
becomes a constant. The change of a single parameter, keeping the other
fixed, is obtained including a Metropolis step in the algorithm, whose
acceptance probability is:
\begin{equation}
\begin{gathered}
\mathcal{P}(\beta_{\text{\textit{old}}}\to\beta_{\text{\textit{new}}}) = 
\min(1, \, e^{-\Delta \beta E_L + \Delta F_L}) \, ,\\
\mathcal{P}(\theta_{\text{\textit{old}}}\to\theta_{\text{\textit{new}}}) = 
\min(1, \, e^{\Delta \theta Q_L + \Delta F_L}) \, .
\end{gathered}
\end{equation}
The free energy as a function of $\beta_L$ and $\theta_L$ is
estimated by a numerical integration of $\braket{E_L}$ and 
$\braket{Q_L}$~\cite{vicari_simul_temp_cpn}, which can be easily measured on the lattice:
\begin{equation}
\begin{gathered}
\frac{\partial F_L}{\partial \beta_L} = \braket{E_L} \equiv U_L, \\
\frac{\partial F_L}{\partial \theta_L}  = -\braket{Q_L} = - k_{1} 
\, .
\end{gathered}
\end{equation}
Note that the free energies are needed only up to a global irrelevant
constants, so free energy derivatives contain all the relevant information. To
that purpose we ran some preliminary simulations to measure the $\theta_L$ and
$\beta_L$-dependence of $U_L$ and $k_{1}$ and estimate the value of
$F(\beta_L,\theta_L)$ in a chosen interval of the parameters. 

We fixed the interval of values of $\beta_L$ and $\theta_L$ to be used in the
simulated tempering by using the following criteria:
\begin{itemize}
\item $\beta_{\text{\textit{min}}}$ is chosen in a region where the slowing
down is not significant, $\beta_{\text{\textit{max}}}$ is chosen according to
how close to the continuum limit one wants to arrive, the spacing $\delta
\beta$ among intermediate $\beta_L$ values is chosen so that there is a
reasonable overlap between probability distributions of $E_L$ at different
couplings (a necessary condition to get a reasonable Metropolis acceptance).
We a posteriori verified that the simplest choice of a constant
$\delta\beta$ is sufficient to obtain a reasonably uniform acceptance probability; 
\item the choice of the $\theta$ angles does not affect the slowing
down of topological modes, so $\theta_{\text{\textit{min}}}$,
$\theta_{\text{\textit{max}}}$ and $\delta \theta$ were chosen to have a
reasonably large interval and enough points for the imaginary-$\theta$ global fit described in
subsection \ref{def_imm_theta_method}.
\end{itemize}

From a practical point of view, simulated tempering has been implemented
by trying, after each sweep, a change in either $\beta_L$ or $\theta_L$ (with
equal probability) to one of their nearest-neighbor values (with equal
probability). Notice that simulated tempering has not been used for all values
of $N$; in particular for $N < 15$, for which the improvement is marginal, only
standard local updating sweeps have been used.

\subsection{Cooling procedure}\label{subsection_cooling} 

To dampen the effects
of the ultraviolet (UV) fluctuations in the measure of the lattice topological charge, we
adopted a standard smoothing algorithm.  Various lattice studies have shown the
equivalence of different procedures, like cooling~\cite{Berg:1981nw}, the
gradient flow~\cite{Luscher:2010iy}, or smearing, once they are appropriately
matched to each other~\cite{Alles:2000sc, cooling_vs_gradient_flow,
Alexandrou:2015yba}.  In this study we chose cooling, because of its relative
simplicity and speed.

Each cooling step consists in a sweep of the lattice in which every site/link
variable is aligned to its local force:
\begin{equation*}
\phi_{\text{\textit{cooled}}}=\frac{F_\phi}{\vert F_\phi \vert}.
\end{equation*}
This update corresponds to a local minimization of the action and {it typically} does not
alter the topological content of the configuration.  For the cooling procedure
the minimized action does not need to be the same governing the field dynamics.
Therefore, to speed up the cooling process, we chose the unimproved $\theta =
0$ action obtained by setting $c_1=1$ and $c_2=0$ in
Eq.~(\ref{Symanzik_improved_lattice_action_cpn}).

\begin{figure}[htb!]
\hspace*{-0.4cm}
\includegraphics[scale=0.45]{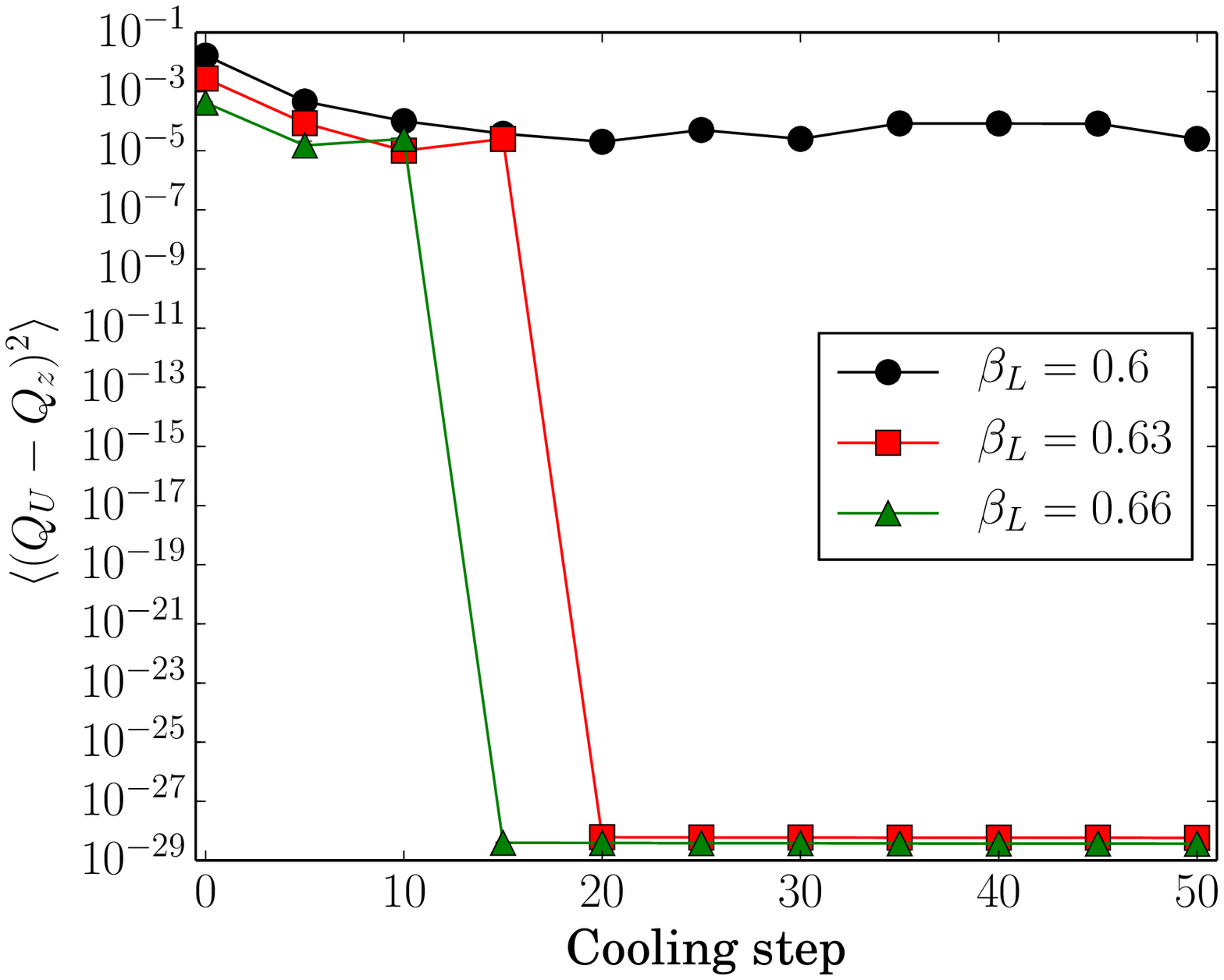}
\hspace*{-0.4cm}
\includegraphics[scale=0.45]{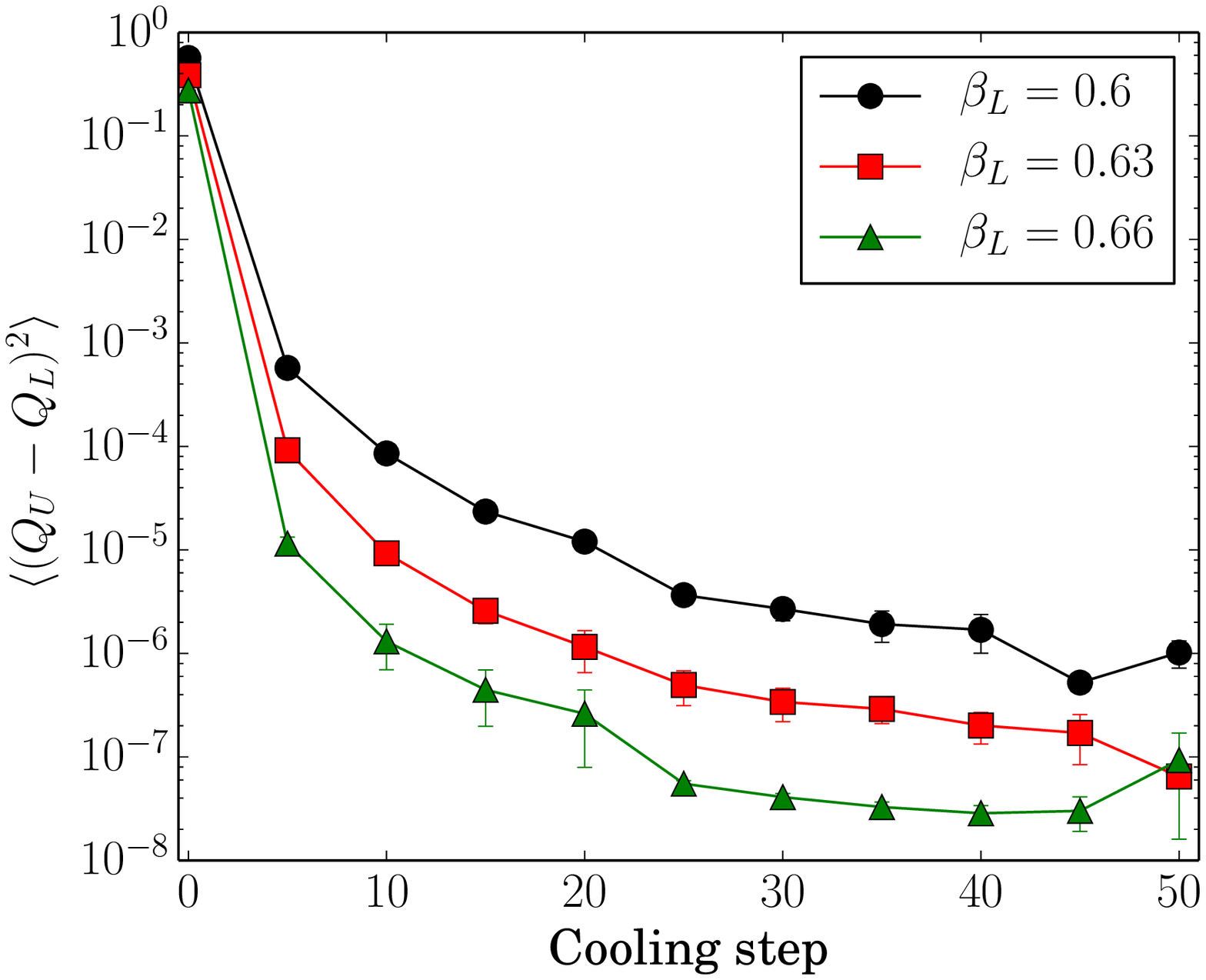}
\caption{Average squared difference between different definitions of the
topological charge as a function of the number of cooling step.
Data have been collected for $N = 21$, $\theta_L = 0$ and
different values of $\beta_L$.  Typical error bars are smaller than the symbol
size.}
\label{fig_charge_cooling}
\end{figure}

Since the various charge discretizations discussed in subsection
\ref{subsection_lattice_action} differ at finite lattice spacing, we checked
that all of them agree after a certain amount of cooling steps. In
Fig.~\ref{fig_charge_cooling} we show an example for $N=21$, $\theta_L=0$ and
some values of $\beta_L$. It turns out that after $\sim 10$ cooling steps all
the definitions agree regardless of $\beta_L$. Moreover, as we move towards the
continuum limit, the discrepancy between the various definitions is less and
less visible, as expected.

\section{Numerical Results}
\label{section_results}

\subsection{Simulations set-up}

\begin{figure}[!htb]
\hspace*{-0.4cm}
\includegraphics[scale=0.45]{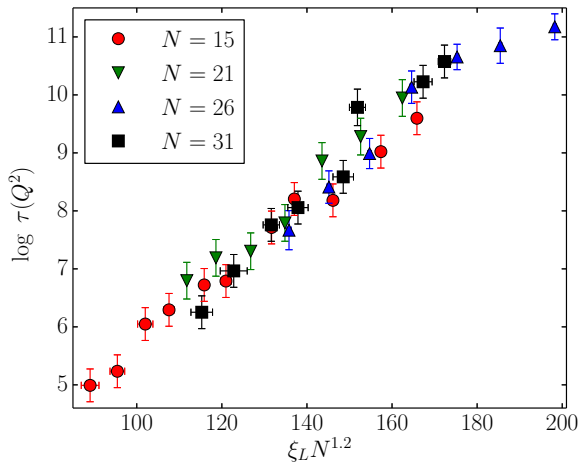}
\caption{Behavior of the autocorrelation time in number of sweeps $\tau(Q^2)$
as a function of the empirical scaling variable $\xi_L N^{1.2}$.}
\label{collasso_tau_xi_N}
\end{figure}

We performed several sets of simulations for
each value of $N$. To start with, we performed numerical simulations at fixed
values of $\beta_L$ and $\theta_L$ in order to determine the free energy
$F_L(\beta_L,\theta_L)$ entering simulated tempering. From some of these runs
we have extracted the autocorrelation time (in number of sweeps) $\tau$ of
$Q^2$ for various values of $\xi_L (\beta_L)$ and $N$ in the case of the
standard local algorithm. As it is well known, critical slowing down
implies an exponential dependence of $\tau(Q^2)$ both on $\xi_L$ and on
$N$~\cite{critical_slowing_down_review, fighting_slowing_down_cpn}. It is
interesting to observe that such a dependence seems to take the form of
a universal function of the scaling variable $\xi_L N^\alpha$, as one can see
from Fig.~\ref{collasso_tau_xi_N}. The coefficient
$\alpha$, fixed empirically to obtain a reasonable collapse of
data points, turns out to be $\alpha = 1.2(1)$.

\begin{figure}[!htb]
\hspace*{-0.4cm}
\includegraphics[scale=0.45]{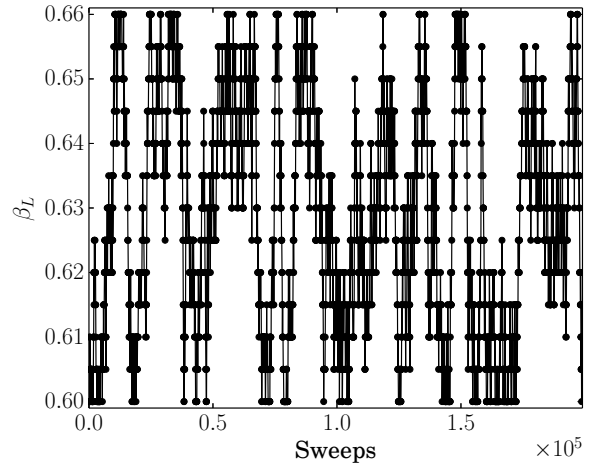}
\caption{Time evolution of $\beta_L$ for $N=21$ during a
time window which is $\sim 0.1 \%$ of the whole accumulated statistics. On the
horizontal axis the time is reported in number of sweeps.}
\label{fig:betathetastories}
\end{figure}

After this preliminary set-up, we performed simulations exploiting simulated
tempering both in $\beta_L$ and in $\theta_L$, as explained above. The total
accumulated statistics for each value of $N$ has been of the order of $10^{9}$
- $10^{10}$ total updating sweeps, typically divided
in $O(10^3)$ independent runs. As an example, in
Fig.~\ref{fig:betathetastories} we report the Monte Carlo evolution of $\beta_L$ during
a simulation performed for $N = 21$.

The simulated tempering provides a gain of a factor up to 4 (for the largest
explored $N$) in terms of the autocorrelation time (measured in number
of sweeps) $\tau(Q^2)$ for the highest $\beta_L$ (which is the one mostly
affected by the freezing). This result is obtained discarding all the
intermediate values of $\beta_L$ and keeping into account the machine time
needed to generate them. The use of all the intermediate couplings to extract
the continuum limit provides a further gain with respect to the case of independent simulations, although
correlations between measures at different couplings introduced by the
algorithm may, in principle, reduce this second advantage.

A number of independent dedicated simulations at
fixed $\theta_L = 0$, with statistics of the order of $10^{9}$ total sweeps for
each $N$, were performed in order to obtain precise measurements of $\xi_L$
needed to accurately determine $\xi_L^2\chi$.
A summary of our simulation
parameters and statistics is reported in Table~\ref{table_info_simulations}.

\begin{table}[!htb]
\begin{center}
\begin{tabular}{ | c | c | c | c | c | c | c | c | c | c | c | } 
\hline
& & & & & & & & & &\\[-1em]
$N$ & Alg. & $\beta_{\text{\textit{min}}}$ & $\beta_{\text{\textit{max}}}$ & $\delta \beta$ & $\theta_{\text{\textit{min}}}$ & $\theta_{\text{\textit{max}}}$ & $\delta \theta$ & Sweeps & $L$ & $\dfrac{L}{\xi^{\text{\textit{max}}}_L}$ \\
& & & & & & & & & &\\[-1em]
\hline
& & & & & & & & & &\\[-1em]
9  & loc.   & 0.8  & 0.9  & 0.02  & 0 & 6 & 0.1 & $2 \cdot 10^{10}$ & 192 & 16\\
11 & loc.   & 0.65 & 0.79 & 0.02  & 0 & 6 & 0.1 & $2 \cdot 10^{10}$ & 100 & 14\\
13 & loc.   & 0.64 & 0.78 & 0.02  & 0 & 6 & 0.1 & $2 \cdot 10^{10}$ & 100 & 14\\
15 & simul. & 0.65 & 0.75 & 0.005 & 0 & 6 & 0.1 & $5 \cdot 10^9$    & 90  & 14\\
21 & simul. & 0.6  & 0.66 & 0.005 & 0 & 5 & 0.1 & $7 \cdot 10^9$    & 72  & 16\\
26 & simul. & 0.6  & 0.64 & 0.005 & 0 & 6 & 0.1 & $7 \cdot 10^9$    & 72  & 18\\
31 & simul. & 0.52 & 0.58 & 0.005 & 0 & 6 & 0.1 & $8 \cdot 10^9$    & 72  & 20\\
\hline
\end{tabular}
\end{center}
\caption{Summary of the simulation parameters adopted for all values of $N$. We
also report the total accumulated statistics, in terms of total sweeps. Measures were performed every 50 sweeps.}
\label{table_info_simulations}
\end{table}

\subsection{Analysis procedure and results}

\begin{figure}[!htb]
\centering
\hspace*{-0.6cm}
\includegraphics[scale=0.45]{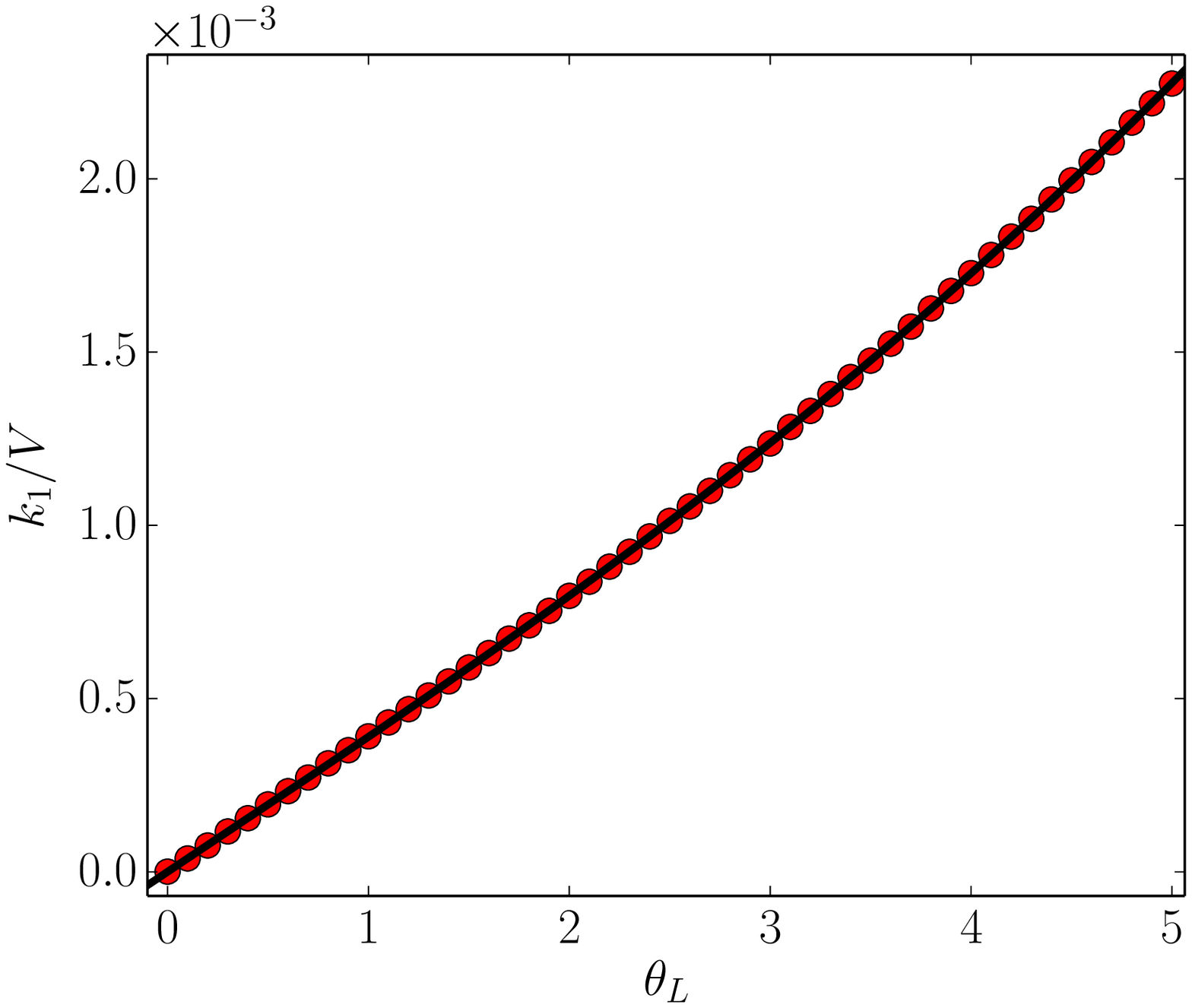}
\hspace*{-0.6cm}
\includegraphics[scale=0.45]{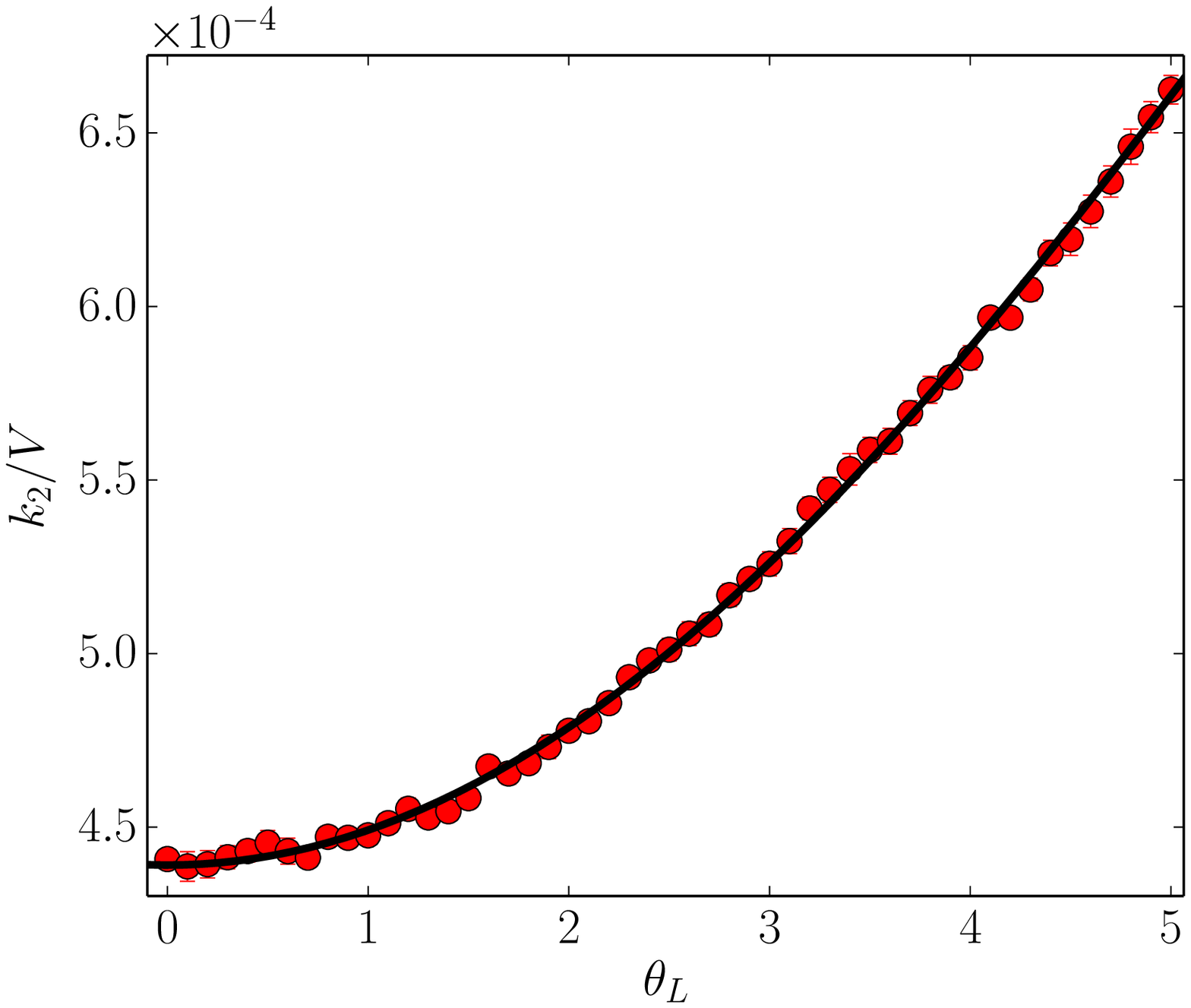}
\hspace*{-0.6cm}
\includegraphics[scale=0.45]{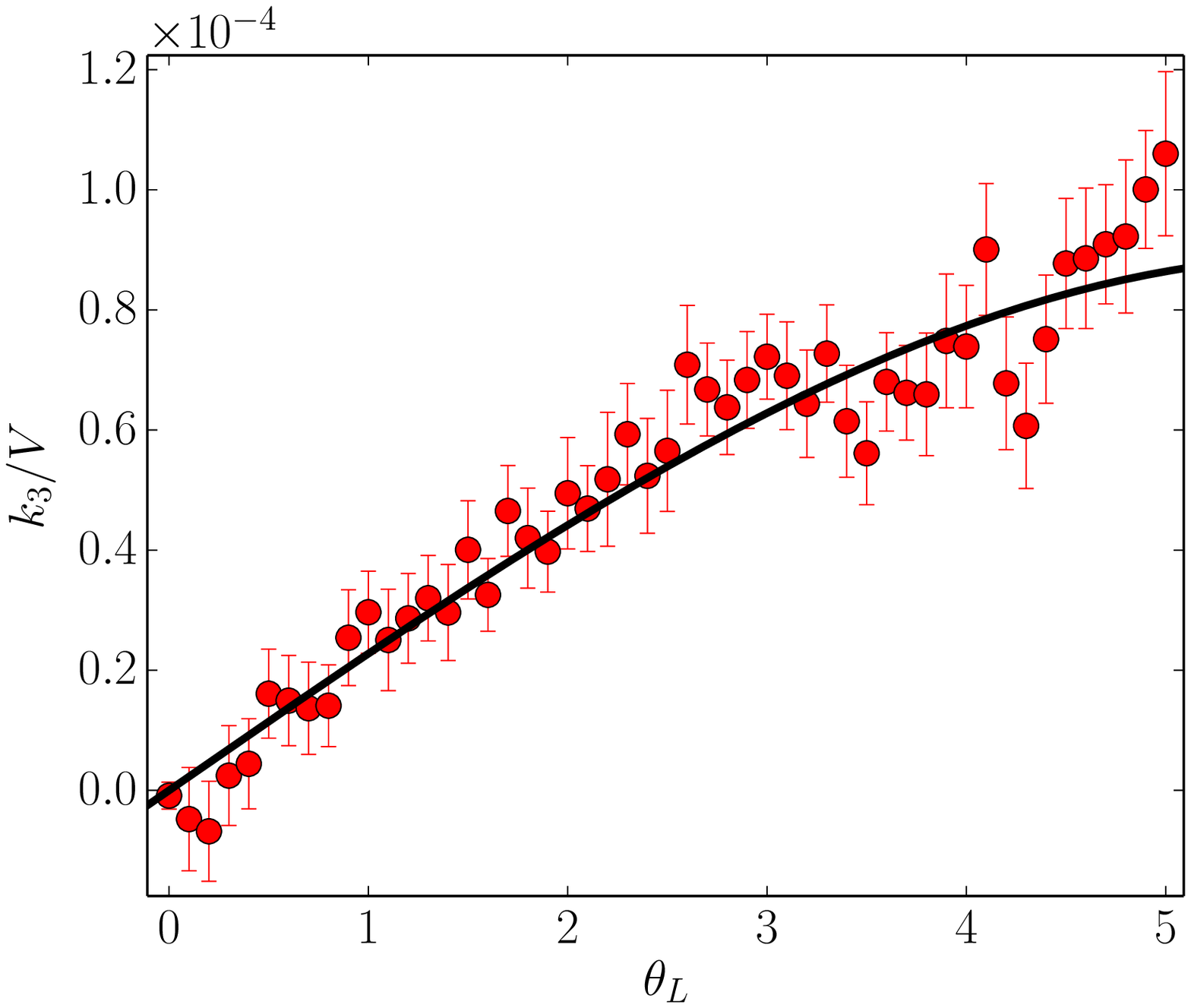}
\caption{Example of the $\theta_L$-dependence of the first three cumulants, along
with best fits obtained expanding $f(\theta)$ up to sixth order. 
Data refers to the case $N = 21$, $\beta_L = 0.66$ on a square lattice with $L = 72$.
In this figure correlations among measures at different $\theta_L$ introduced by the simulated
tempering are ignored.
}
\label{fig_imm_theta_fit_example}
\end{figure}

The simulated tempering provides samples containing measures of $Q_U$ at
different values of $\beta_L$ and $\theta_L$.  Our idea is to fully exploit the
information contained in these samples proceeding as follows:

\emph{i)} we compute the cumulants of $Q_U$ and we measure $\chi$, $b_2$ and
$b_4$ for each value of $\beta_L$ using a global fit to
Eqs.~(\ref{imm_theta_fit}) (an example is shown in
Fig.~\ref{fig_imm_theta_fit_example});

\begin{figure}[htb!]
\centering
\hspace*{-0.9cm}
\includegraphics[scale=0.45]{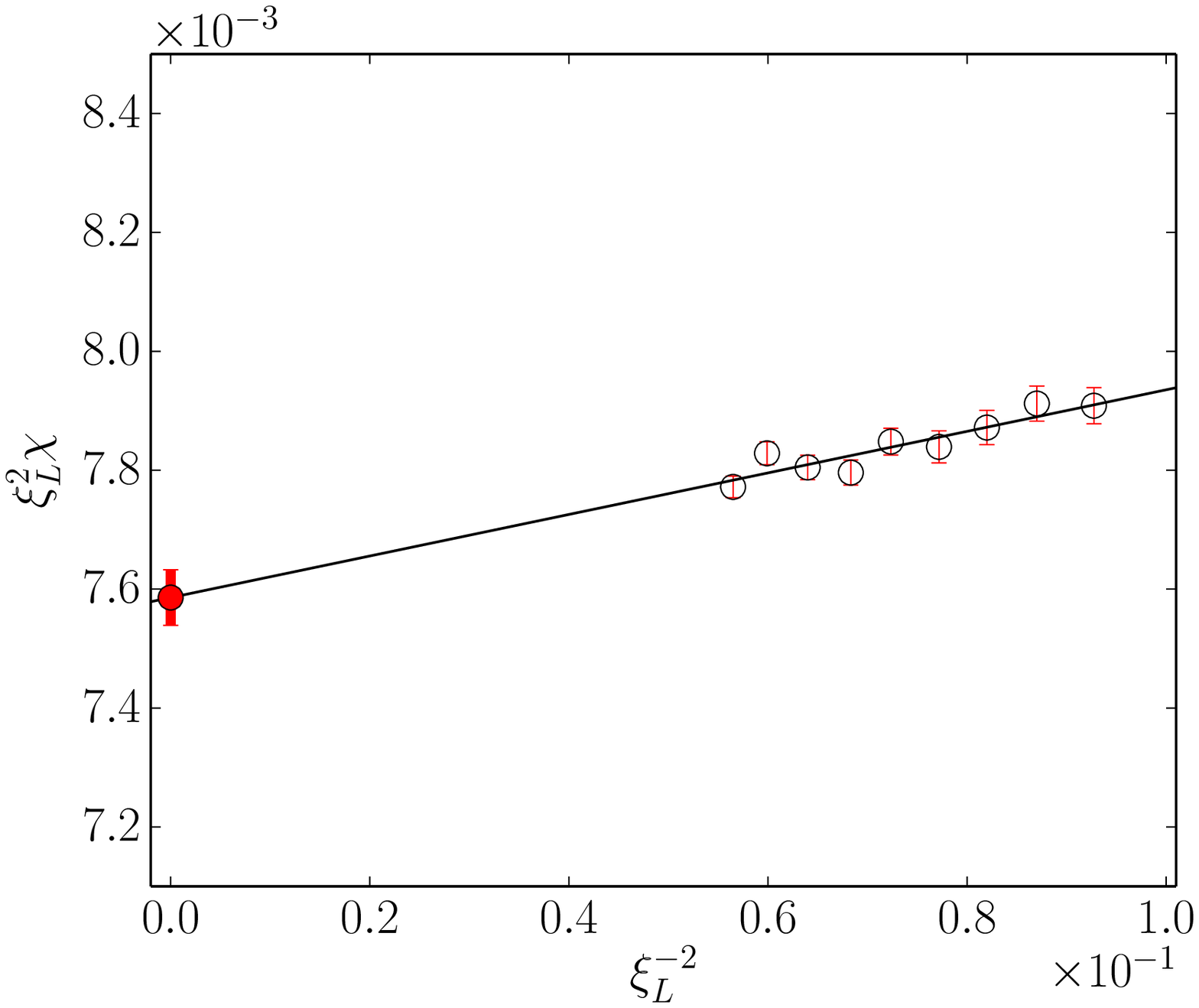}
\hspace*{-0.9cm}
\includegraphics[scale=0.45]{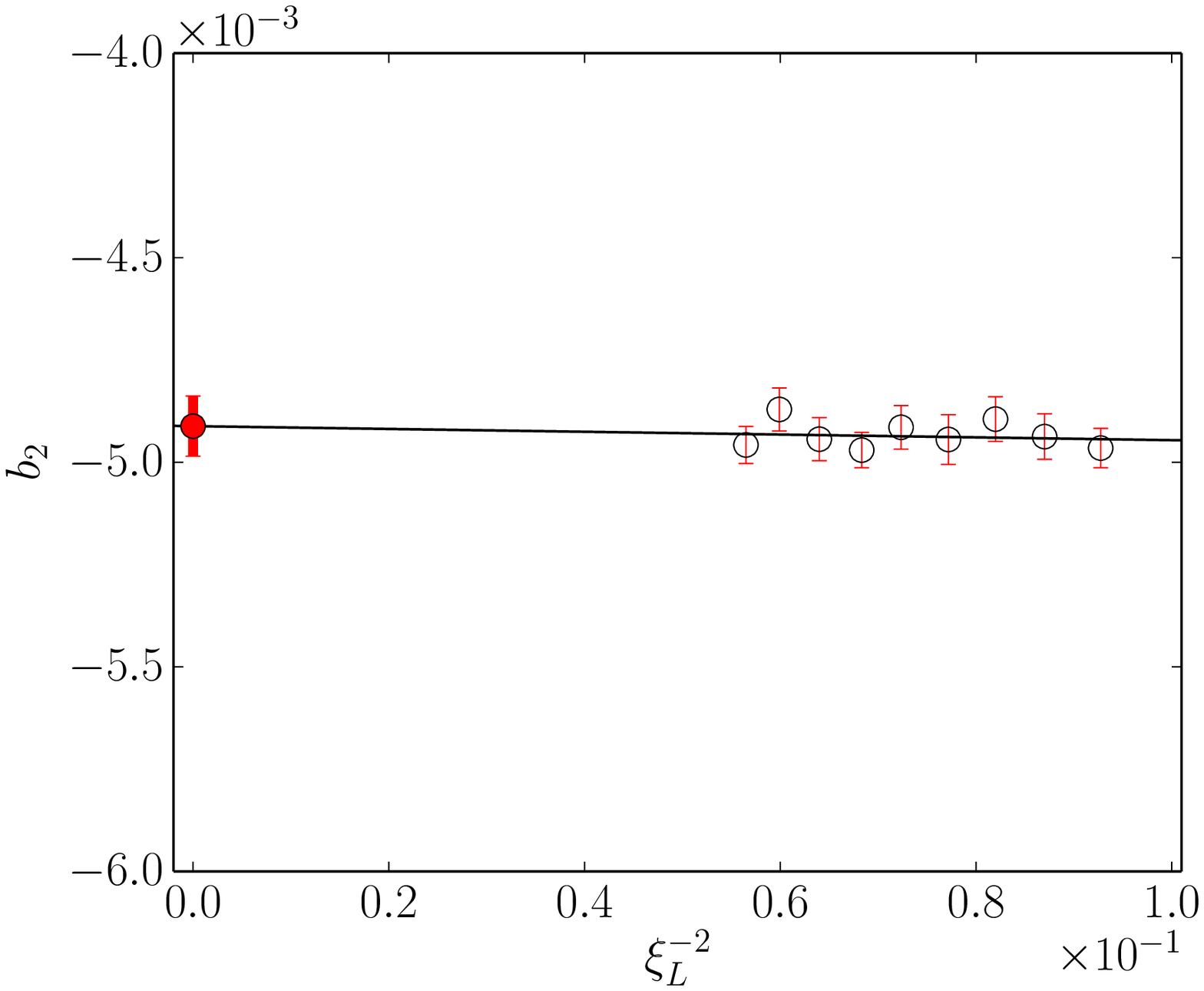}
\hspace*{-0.9cm}
\includegraphics[scale=0.45]{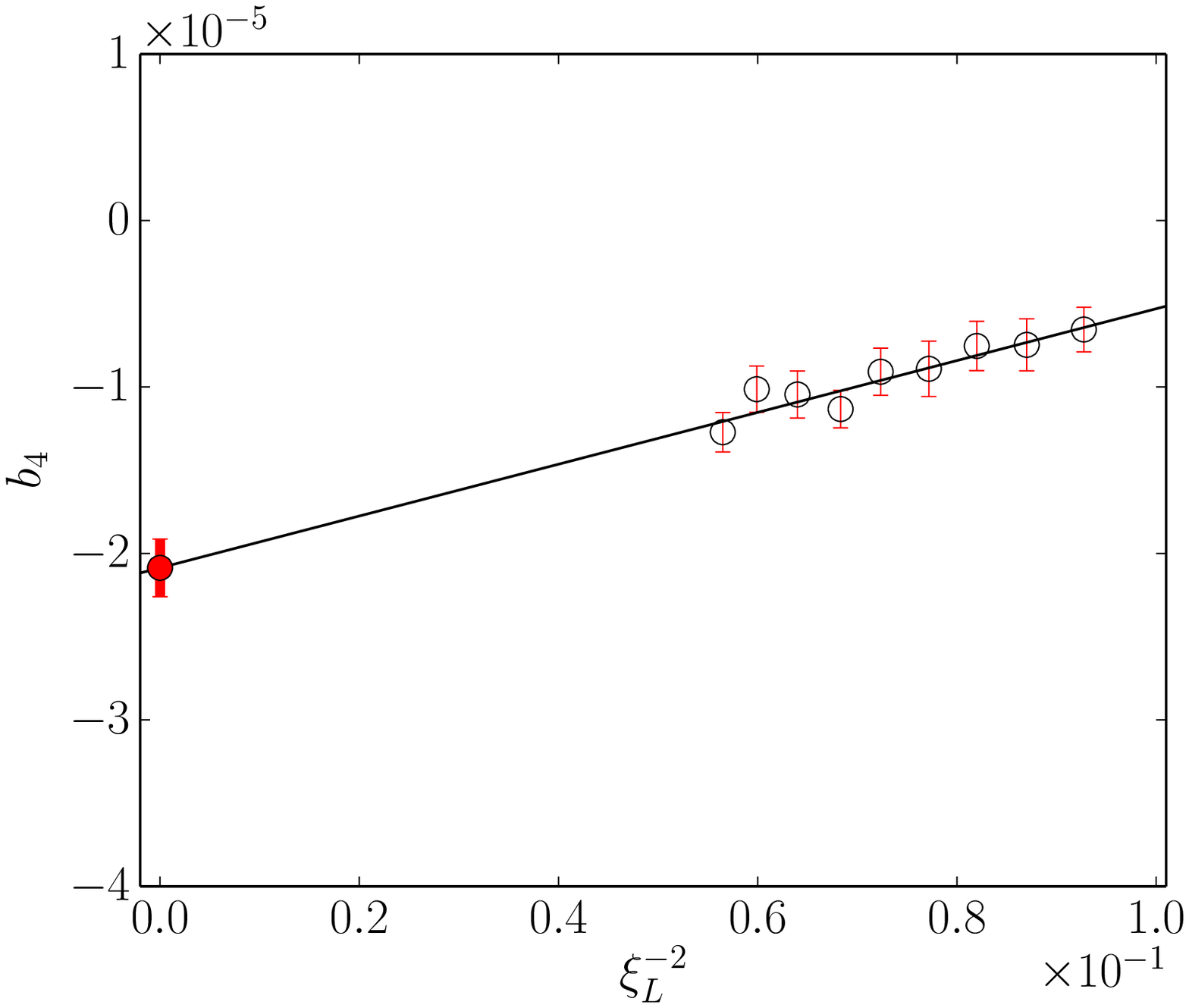}
\caption{Extrapolation to the continuum limit for $N = 21$ using the complete
data sets and ignoring correlations among measures at different $\beta_L$
introduced by the simulated tempering. The extrapolations are plotted as filled
points. Starting from above, we show results for $\xi^2 \chi$, $b_2$ and $b_4$,
whose continuum values are respectively $\xi^2 \chi =
7.59(5) \cdot 10^{-3}$, $b_2=-4.9(1)\cdot10^{-3}$ and
$b_4=-2.1(2)\cdot10^{-5}$. Fits are stable varying the fit range and the
reduced $\tilde\chi^2$ (with 7 degrees of freedom) is 0.9, 0.5 and 0.4
respectively. Comparing these results with the ones reported in Table
\ref{table_results} we observe that correlations affect the final {estimate} of
the error just for $b_2$ and $b_4$.}
\label{fig_continuum_limit_example}
\end{figure}

\begin{figure}
\centering
\hspace*{-0.9cm}
\includegraphics[scale=0.45]{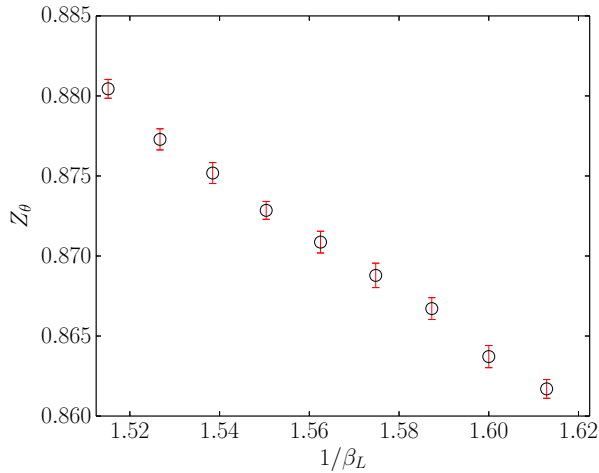}
\caption{{$Z_\theta$ results obtained for $N=21$. Data are plotted
against $1/\beta_L$ since deviations of $Z_{\theta}$ from 1 can be expanded in
powers of $1/\beta_L$}~\cite{renormalization_lattice_top_charge_1}. {Note
that the values of $Z_{\theta}$ obtained in this work are much closer to $1$
than those obtained in $SU(N_c)$ Yang-Mills theories, see, e.g.,}
Ref.~\cite{SU(N)_large_N_limit}.}
\label{Z_theta_versus_1_over_beta_L}
\end{figure}

\emph{ii)} we exploit the determinations at all $\beta_L$ values to obtain the
continuum limit extrapolation of $\xi^2 \chi$, $b_2$ and $b_4$ for a given
value of $N$, through a best fit to Eq.~(\ref{continuum_limit_scaling_xi_L})
(an example of this extrapolation is reported in
Fig.~\ref{fig_continuum_limit_example} for $N = 21$, {with the corresponding $Z_{\theta}$ values 
shown in Fig.~\ref{Z_theta_versus_1_over_beta_L}}).

The simulated tempering introduces correlations among measures at different
$\theta_L$ and $\beta_L$ values that come from the same Markov chain.
Thus, it is of primary importance to carefully take these correlations into
account to get a correct estimate of the statistical
errors.  To do so, we applied a bootstrap analysis to our data set,
treating as a single independent extraction a whole set of measures coming from
one of the $O(10^3)$ independent simulated tempering runs performed for each
$N$.  Of course for runs at $N \leq 13$, for which simulated tempering
was not adopted, this analysis was not required and we relied on a simple
binned jackknife analysis at each fixed $\beta_L$ and $\theta_L$.

Several sources of systematic error were checked. For the imaginary-$\theta$
fit we verified that the fit parameters, as well as the
reduced $\tilde \chi^2$, were stable varying the number of cumulants and the
$\theta$-interval used in the fit. Concerning the truncation of the fit
functions, stopping to the $b_4$ terms in each cumulant was always sufficient
to get stable results and reduced $\tilde \chi^2$ of order 1, while the
introduction of higher-order terms
in $\theta$ in the fit did not change results within errors. As for the
extrapolation to the continuum limit, fitting with a linear function of
$\xi^{-2}_L$ was always sufficient to get a reduced $\tilde\chi^2$ of order 1.
Furthermore, we checked that both the continuum limit and the reduced
$\tilde{\chi}^2$ were stable when varying the fit range.
A final remark about finite-size corrections:
we checked that in all our simulations $L/\xi_L > 10$
and $(L/\xi_L)^2\gg N$ (see
Ref.~\cite{Aguado:2010ex} for more details on this condition), so they are
under control.

\FloatBarrier

A summary of our continuum extrapolated results is reported in
Table~\ref{table_results}, where we also report some results from other
recent studies~\cite{fighting_slowing_down_cpn, csd_cpn_sanfo_unpublished}.

\begin{table}[htb!]
\begin{center}
\begin{tabular}{ | c || c | c | c | } 
\hline
& & & \\[-1em]
$N$ & $\xi^2 \chi \cdot 10^3$ & $b_2 \cdot 10^3$ & $b_4 \cdot 10^5$ \\
\hline
& & & \\[-1em]
9    & 20.00(15)  & -13.90(13)  & 2.04(18) \\
10*  & 17.37(8)   & -           & -        \\
10** & 17.19(10)  & -           & -        \\
11   & 15.24(12)  & -10.70(40)  & 2.30(50)   \\
13   & 12.618(88) & -9.10(32)   & -0.62(52)  \\
15   & 10.87(11)  & -6.69(26)   & 0.52(57)   \\
21*  & 7.67(5)    & -           & -        \\
21   & 7.588(49)  & -4.98(53)   & -1.7(1.2)    \\
26   & 6.141(53)  & -3.01(39)   & -0.33(53)  \\
31   & 5.028(60)  & -2.31(22)   & -0.097(0.28)  \\
41*  & 3.91(2)    & -           & -        \\
\hline
\end{tabular}
\end{center}
\caption{In this table we report continuum
extrapolated results for different values of $N$.  We
report also some results obtained in other recent studies. In particular,
results marked {by} * and ** are taken from
Refs.~\cite{fighting_slowing_down_cpn} and \cite{csd_cpn_sanfo_unpublished},
respectively; in these works only $\xi^2 \chi$ is
reported.}
\label{table_results}
\end{table}

It is important to stress the good agreement between continuum extrapolated
values obtained by studies adopting different discretizations: 
for $N = 21$ our result for $\xi^2\chi$ (Symanzik improved discretization)
agrees within one standard deviation with that reported in
Ref.~\cite{fighting_slowing_down_cpn} (standard discretization).  This
represents a consistency check that the one percent precision level achieved in
both cases is solid, i.e. it correctly takes into account the possible sources
of statistical and systematic uncertainty.

As for the $b_4$ coefficient, we notice that our present numerical precision
only allows to set upper bounds after the continuum extrapolation is taken,
with the only exceptions of $N=9$ and $11$, where a reliable non-zero
determination has been obtained.

\section{Discussion and the large-$N$ limit}
\label{results_chi}

\subsection{Numerical large-$N$ limit for $\xi^2 \chi$ and $b_2$}

The main purpose of this section is to make use of our results and of results
from other recent studies~\cite{fighting_slowing_down_cpn,
csd_cpn_sanfo_unpublished}, all summarized in Table~\ref{table_results}, to
investigate the large-$N$ behavior of the free energy coefficients and compare
it with analytic predictions. 

Let us start from the topological susceptibility: its analytic large-$N$
prediction has been summarized in Eq.~(\ref{large_N}). If one tries to
numerically check the leading order term and fits data with $N \geq 26$
using $\xi^2 \chi = e_1/N$, one obtains $e_1 = 0.1600(8)$
with $\tilde\chi^2=2.86/2$, which is in good agreement with the analytic result
$1/(2\pi)\simeq 0.1592$. This result is stable both under the change of the fit
range and under the inclusion of a further quadratic
term in $1/N$.

\begin{figure}[!htb]
\centering
\hspace*{-0.9cm}
\includegraphics[scale=0.45]{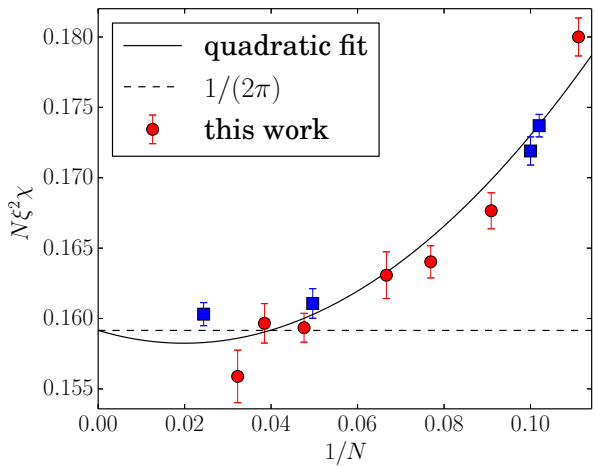}
\caption{{Large-$N$ behavior of} $N\xi^2 \chi$. {We report also} the values obtained in
Refs.~\cite{fighting_slowing_down_cpn} and \cite{csd_cpn_sanfo_unpublished}. 
}
\label{fig_grande_N_susc_top}
\end{figure}

The good agreement of the leading $1/N$ term with analytic predictions was
already noted in previous studies where, however, some tension emerged when
comparing with the next-to-leading analytic computation, $e_2 \simeq -0.0606$.
This seeming discrepancy is clearly visible from
Fig.~\ref{fig_grande_N_susc_top}, where $N \xi^2 \chi$ is reported together
with the asymptotic result $1/(2 \pi)$ for this quantity: deviations from the
asymptotic value are typically positive. On the other hand, if one tries to fit
all data according to $N \xi^2 \chi = 1/(2 \pi) + e_2/N$, one obtains
the positive value $e_2 = 0.12(1)$, but with $\tilde\chi^2 =
130/10$, meaning that results cannot be accounted for by just the
next-to-leading contribution: higher order terms are needed.  As one can see
from Table \ref{table_large_N_chi_versus_1_over_N}, this is true even if the
smallest values of $N$ are discarded from the fit: the $\tilde\chi^2$ is still
large and the fitted value of $e_2$ is not stable.

Therefore, we tried to include a $e_3/N^3$ correction in $\xi^2 \chi$: as
one can see from Table \ref{table_large_N_chi_versus_1_over_N}, in this case
$e_2$ turns out to be negative and actually compatible with the analytic
prediction (even if with large error bars), with marginally
acceptable values of $\tilde \chi^2$ when the smallest ($N \leq
10$) values of $N$ are discarded.  One can even fix $e_2$ to its
analytic prediction, without changing the quality of the fit and maintaining a
stable prediction for $e_3$ in the range $1.5 - 2$. 

\begin{table}[!htb]
\begin{center}
\begin{tabular}{ | c | c | c | c | c | c | } 
\hline
& & & & & \\[-1em]
$N_{min}$ & $e_1$ & $e_2$ & $e_3$ & $\tilde \chi^2$ & dof\\
\hline
& & & & & \\[-1em]
26 & $1/2\pi$ & 0.011(36)   & & 2.47 & 2  \\
21 & "        & 0.018(16) & & 1.62 & 4  \\
15 & "        & 0.027(15) & & 1.71 & 5  \\
13 & "        & 0.040(13) & & 2.07 & 6  \\
11 & "        & 0.056(14) & & 3.23 & 7  \\
10 & "        & 0.105(16) & & 10.6 & 9  \\
9  & "        & 0.116(16) & & 13.5 & 10 \\
\hline
\hline
& & & & & \\[-1em]
21 & " &  0.019(95)   & -0.026(2.2)    & 2.16 & 3 \\
15 & " & -0.023(61)   & 1.0(1.2)     & 1.82 & 4 \\
13 & " & -0.028(38) & 1.17(63)   & 1.46 & 5 \\
11 & " & -0.040(29) & 1.41(41) & 1.28 & 6 \\
10 & " & -0.077(29) & 2.09(33)   & 1.95 & 8 \\
9  & " & -0.092(30) & 2.30(33)   & 2.30 & 9 \\
\hline
\hline
& & & & & \\[-1em]
26 & " & -0.0606 & 2.1(1.3) & 3.2 & 2 \\
21 & " & "       & 1.79(42) & 2.0 & 4 \\
15 & " & "       & 1.79(29) & 1.6 & 5 \\
13 & " & "       & 1.68(17) & 1.4 & 6 \\
11 & " & "       & 1.69(12) & 1.2 & 7 \\
10 & " & "       & 1.914(72)  & 1.8 & 9 \\
9  & " & "       & 1.977(74)  & 2.3 & 10\\
\hline
\end{tabular}
\end{center}
\caption{In this table we report systematics for the 
{determination of the large-$N$ behavior} 
of $\xi^2 \chi$ using the fit function $N\xi^2\chi = e_1 +
e_2/N + e_3/N^2$. Blank spaces mean that the corresponding coefficient
was set to 0 in the fit procedure while numerical values with no error mean
that the corresponding coefficient was fixed to that value.}
\label{table_large_N_chi_versus_1_over_N}
\end{table}

Let us summarize what, in our opinion, one can conclude from this analysis:
presently available results are actually consistent with the analytic
next-to-leading large-$N$ prediction for $\xi^2 \chi$, once a further term in
the $1/N$ expansion is taken into account.  However, an independent and
numerically stable determination of $e_2$ is presently not possible: this is
due to the fact that $e_2$ and $e_3$ turn out be of opposite signs, with $e_3$
at least one order of magnitude larger than $e_2$, so that for the largest
available values of $N$ their contributions still cancel each other, while for
the smallest values ($N \lesssim 10$) one cannot exclude that further terms
could come into play. In order to make a further improvement, in the future one
should try to substantially increase the precision of present data
and/or to push the investigation towards even larger values of $N$.

\begin{figure}[!htb]
\centering
\includegraphics[scale=0.45]{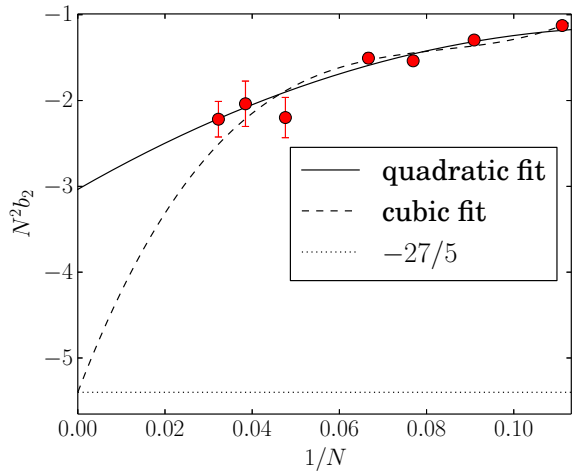}
\caption{{Large-$N$ behavior of $N^2 b_2$. A quadratic and a cubic fit are also
shown and in the cubic case the constant term was fixed to the theoretical value.}
} \label{fig_grande_N_b_2}
\end{figure}

We now pass to the analysis of $b_2$. In
this case, apart from some preliminary study~\cite{tesi_manca}, no 
results can be found in the literature, therefore this
work represents the first numerical test of the large-$N$ scaling of
this coefficient. The leading order contribution is expected to be of order
$1/N^2$, see Eq.~(\ref{b2largen}); for this reason, we have considered the
quantity $N^2 b_2$, which is expected to have a finite large-$N$ limit and is
plotted in Fig.~\ref{fig_grande_N_b_2}.

Since results clearly point to a finite value of $N^2 b_2$ as $1/N \to 0$ (see
Fig.~\ref{fig_grande_N_b_2}), we tried to fit data {according} 
to the following function:
\begin{equation}
N^2 b_2 = \bar{b}_{2} + k_1\frac{1}{N}+k_2\frac{1}{N^2}\ ;
\label{fitb2}
\end{equation}
fitting in the whole range we obtain:
\begin{equation}
\bar{b}_2 = -3.0(5), \quad \tilde \chi^2 =6/4.
\end{equation}
This result is stable if we change the fit range and we obtain compatible
results fixing $k_2 = 0$ and fitting in a
narrower range, however it is not compatible with the analytic prediction
in Eq.~(\ref{b2largen}) $\bar{b}_2=-27/5=-5.4$. Also in this case, a
possible slow convergence to the large-$N$ asymptotic regime could explain the
disagreement: indeed, if in Eq.~(\ref{fitb2}) we fix $\bar{b}_2=-27/5$, we are
still able to obtain a good fit, with $\tilde \chi^2 =6/4$, by allowing for
just another $k_3/N^3$ term in the expansion. Both best fits are reported in
Fig.~\ref{fig_grande_N_b_2}.  As one can appreciate from
this figure, a definite assessment of this issue will be
possible with more precise data at the largest explored values of $N$, or once
results will be available for $1/N \lesssim 0.02$, i.e.~$N \gtrsim 50$.

Results for the $b_4$ coefficient will be discussed separately 
in Sec.~\ref{subsection_b_4}.

\begin{table}[!htb]
\begin{center}
\begin{tabular}{ | c | c | c | c | c | c | c | } 
\hline
& & & & & &\\[-1em]
$N_{min}$ & $\bar{b}_2$ & $k_1$ & $k_2$ & $k_3$ & $\tilde \chi^2$ & dof \\
\hline
26 & -2.149(87) & & & & 0.29 & 1\\
21 & -2.165(53) & & & & 0.16 & 2\\
\hline
\hline
& & & & & & \\[-1em]
21 & -2.20(45) & 0.85 $\pm$ 11 & & & 0.31 & 1 \\
15 & -3.00(28) & 22.3(4.4)         & & & 0.75 & 2 \\
13 & -2.54(39) & 13.9(5.6)         & & & 2.34 & 3 \\
11 & -2.51(24) & 13.4(3.0)         & & & 1.76 & 4 \\
9  & -2.31(13) & 10.6(1.2)         & & & 1.74 & 5 \\
\hline
\hline
& & & & & & \\[-1em]
13 & -4.1(1.0) & 70(35) & -480(300) & & 1.54 & 2 \\
11 & -3.04(76) & 29(22) & -112(150) & & 1.99 & 3 \\
9  & -2.92(47) & 25(11) & -82(62)   & & 1.52 & 4 \\
\hline
\hline
& & & & & & \\[-1em]
13 & $-27/5$ & 141(27) & -1700(800) & 6500(6500)  & 2.0 & 2 \\
11 & "     & 154(16) & -2000(450) & 10000(3000) & 1.6 & 3 \\
9  & "     & 132(12) & -1500(270) & 6050(1450)  & 2.3 & 4 \\
\hline
\end{tabular}
\end{center}
\caption{In this table we report systematics for the 
{determination of the large-$N$ behavior}
 of $b_2$ obtained using the fit function $N^2 b_2 = \bar{b}_2 +
k_1/N + k_2/N^2+k_3/N^3$. The convention is the same of Table
\ref{table_large_N_chi_versus_1_over_N}.}
\label{table_b_2_large_N}
\end{table}

\subsection{Trying to match the large-$N$ and the small-$N$ behaviors}

As we have discussed above, numerical results for $\xi^2 \chi$ clearly indicate
that, within a $1/N$ expansion, at least $O(1/N^3)$ contributions are needed
for $N \sim O(10)$, and even higher order terms are likely to give
non-negligible contributions.  Actually, one must take into account that any
$N$-dependence of topological quantities should match the divergence of the
topological susceptibility which is expected~\cite{Jevicki:1977yd,
Forster:1977jv, Fateev:1979dc, Berg:1979uq,Richard:1981bv}, and has been
numerically verified~\cite{Blatter:1995ik, DElia:1995wxi}, for $N = 2$. A naive
way to match the usual large-$N$ expansion and the divergence in $N=2$ would be
to change the expansion parameter from $1/N$ to\footnote{{For the case of the
$3d$ $O(N)$ models, the similar idea that the expansion parameter $1/(N-1)$
could be more effective than the standard $1/N$ was advocated in
Ref.~\cite{Giombi:2013fka} (see also Ref.~\cite{Jevicki:2014mfa} for further
discussion).}} $1/(N-2)$; however, a more educated guess can be tried, based on
how $\chi$ is expected to diverge as $N = 2$ is approached.

That can be done by recalling how the prediction for a divergence comes
out. In the weak coupling regime, the instanton size distribution for
the $CP^{N-1}$ models is given by
\begin{equation}
P(\rho) = \rho^{N-3},
\end{equation}
thus, the average number of instantons and antiinstantons expected 
in a finite box of size $\rho_0$ is approximately given (for $N>2$) by
\begin{equation}
\braket{N_I} = \braket{N_A} = \int_{0}^{\rho_0}\rho^{N-3} d\rho = \frac{\rho_0^{N-2}}{N-2} \propto \frac{1}{N-2}.
\end{equation}
{Assuming, naively, 
that the instanton-antiinstanton gas is {weakly} 
interacting, as in
the Dilute Instanton Gas Approximation (DIGA),
then 
$N_I$ and $N_A$ are both distributed 
according to two independent Poissonian distributions: using
this fact it is simple to show that}
\begin{equation}
\chi \propto \braket{(N_I-N_A)^2} \propto \frac{1}{N-2} \, ,
\end{equation}
{i.e.~the susceptibility is expected to diverge as $1/(N-2)$.}

Of course, our assumption is questionable: 
the properties
and dynamics of the small topological objects dominating 
at small $N$ can be more complex \cite{Diakonov:1999ae,Andersen:2006sf,Lian:2006ky}, {and they are not diluted at all, since their density diverges 
as $N\to 2$.
However,} it is interesting to see if such a prediction is supported by present numerical
data available for the smallest values of $N$. If we fit data with $N < 15$ to
a functional dependence $\xi^2 \chi = a/(N - b)^c$ we obtain $b = 2.14(18)$
with $\tilde \chi^2 = 5.1/3$ if we fix $c = 1$, and $c = 1.013(22)$ with a
similar $\tilde \chi^2$ if we fix $b = 2$, while our available information is
still not enough to reliably fit both $b$ and $c$ at the same time.
Finally, fixing both $b = 2$ and $c = 1$, one still gets $\tilde \chi^2
= 5.6/4$ with $a = 0.1385(5)$, meaning {(when compared
to $1/(2 \pi) \simeq 0.1592$)}
that the simple $1/(N-2)$ behavior could
still describe up to 90\% of the total signal which is observed even for {asymptotically large $N$ values.}

Therefore, we conclude that our present data are perfectly consistent with the
presence of a $1/(N - 2)$ divergence, and that its amplitude is large enough to
influence data for $N \sim O(10)$.  It is clear that the standard $1/N$
expansion has to face the presence of such a divergence, so that its radius of
convergence cannot be larger than $1/2$.  It is interesting to consider if one
can devise a different form of the expansion, in which the presence of the
divergence is explicitly taken into account, in such a
way that the remaining $N$-dependence is more regular. 

{The divergence at $N=2$ can be enforced in the fit in many different ways. We
will use}
\begin{equation}\label{ansatz_N_meno_2_2}
\xi^2 \chi = \frac{1}{N-2}\, f_N\ ,
\end{equation}
{in which $f_N$ is a function with a regular expansion in $1/N$ and we
factored out the divergent contribution at $N=2$. Another obvious possibility
would be to use the functional form $a/(N-2)+g_N$, but this expression is
exactly equivalent to Eq.~\eqref{ansatz_N_meno_2_2} with $f_N=a + (N-2) g_N$.}
Also note that Eq.~\eqref{ansatz_N_meno_2_2} can be {thought of} as a
partial resummation of the $1/N$ series, since:
\begin{equation}\label{N_meno_2_resummation}
\frac{1}{N-2} = 
\frac{1}{N}\sum_{n=0}^{\infty}\left(\frac{2}{N}\right)^n.
\end{equation}

Expanding $f_N$ in powers of $1/N$, one gets
\begin{equation}\label{fit_function_figata_2}
\xi^2\chi=\frac{1}{N-2}\left[e_1^\prime + \frac{e_2^\prime}{N} + \frac{e_3^\prime}{N^2} + O\left(\frac{1}{N^3}\right)\right]
\end{equation}
and {from} the expansion \eqref{N_meno_2_resummation} one finds
$(e^\prime_1)_{\text{\textit{theo}}}=1/(2\pi)$ and
$(e^\prime_2)_{\text{\textit{theo}}}=-2(e_1)_{\text{\textit{theo}}}+(e_2)_{\text{\textit{theo}}}
\simeq -0.379\,.$ Using this parametrization and fixing $e_1^\prime=1/(2\pi)$,
we fitted $(N-2)\xi^2\chi$ versus $1/N$, obtaining the results reported
in Table~\ref{table_large_N_chi_versus_1_over_N_ansatz_N_meno_2}.
We notice that, although difficulties in
measuring $e_2$ still persist, compared to the standard expansion the resulting
fit is more stable, the reduced $\tilde\chi^2$ is marginally smaller and
the scaling window is larger.  This might suggests that indeed the $N-2$
ansatz  in Eq.~(\ref{ansatz_N_meno_2_2}) constitutes an improved
parametrization, compared to the usual one, of the $N$-dependence of $\xi^2
\chi$, since it encodes more relevant physics.

Finally, let us stress that the divergence of $\xi^2 \chi$ does not
imply that also further coefficients of the $\theta$-expansion should diverge
for $N = 2$. For instance, in the naive non-interacting instanton gas
approximation, all coefficients would stay finite even in the presence of a
divergent instanton density. Actually, future studies, focussing on the
small-$N$ side of the problem, could clarify, by determining $b_2$ and other higher order
terms in the $\theta$ expansion, the exact nature and distribution of the
UV-diverging topological objects populating the small-$N$ world.

\begin{table}[htb!]
\begin{center}
\begin{tabular}{ | c | c | c | c | c | c | } 
\hline
& & & & &\\[-1em]
$N_{min}$ & $e_1^\prime$& $e_2^\prime + 1/\pi$ & $e_3^\prime$ & $\tilde \chi^2$ & dof\\
\hline
& & & & &\\[-1em]
26 & $1/2\pi$ & 0.010(34)   & & 2.5 & 2\\
21 & "        & 0.017(15) & & 1.6 & 4\\
15 & "        & 0.024(13)   & & 1.7 & 5\\
13 & "        & 0.036(11)   & & 1.9 & 6\\
11 & "        & 0.049(11)   & & 2.9 & 7\\
\hline
\hline
& & & & & \\[-1em]
13 & " & -0.022(35)   & 0.98(56) & 1.4 & 5\\
11 & " & -0.031(26) & 1.14(36) & 1.2 & 6\\
10 & " & -0.060(25) & 1.66(27) & 1.7 & 8\\
9  & " & -0.071(25) & 1.82(26) & 1.9 & 9\\
\hline
\hline
& & & & & \\[-1em]
15 & " & $-0.0606$ & 1.72(26)  & 1.6 & 5\\
13 & " & "         & 1.57(16)  & 1.5 & 6\\
11 & " & "         & 1.54(10)  & 1.3 & 7\\
10 & " & "         & 1.673(54) & 1.5 & 9\\
9  & " & "         & 1.712(52) & 1.8 & 10\\
\hline
\end{tabular}
\end{center}
\caption{In this table we report systematics for the 
{determination of the large-$N$ behavior}
of $\xi^2 \chi$ using the $N-2$ ansatz in 
Eq.~(\ref{fit_function_figata_2}). Conventions are the same 
{as} in Table \ref{table_large_N_chi_versus_1_over_N}.}
\label{table_large_N_chi_versus_1_over_N_ansatz_N_meno_2}
\end{table}

\begin{figure}[htb!]
\centering
\includegraphics[scale=0.45]{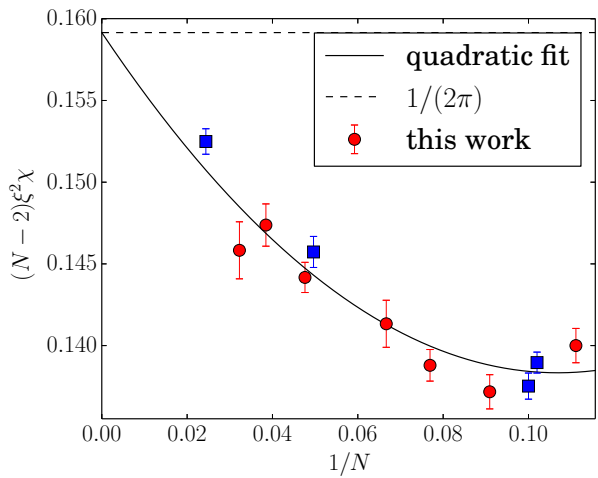}
\caption{Fit to the large-$N$ behavior of the topological susceptibility,
using the $N-2$ ansatz in Eq.~(\ref{fit_function_figata_2}).}
\label{fig_grande_N_susc_top_N_meno_2}
\end{figure}

\subsection{The $O(\theta^6)$ term in $CP^{N-1}$ models and Yang-Mills theories}
\label{subsection_b_4}

We have seen in the previous sections that it is possible to provide precise
estimates of $\chi$ and $b_2$ in the $CP^{N-1}$ models, thus fixing the
$\theta$-dependence of $f(\theta)$ up to $O(\theta^4)$. The numerical
determination of higher orders in $\theta$ is however extremely challenging,
since higher orders are suppressed by large powers of $1/N$ and they are thus
very small. In this section we will present our results for $b_4$, i.e.  the
coefficient fixing the $\theta^6$-dependence of $f(\theta)$.
 
\begin{figure}[htb!]
\centering
\includegraphics[scale=0.45]{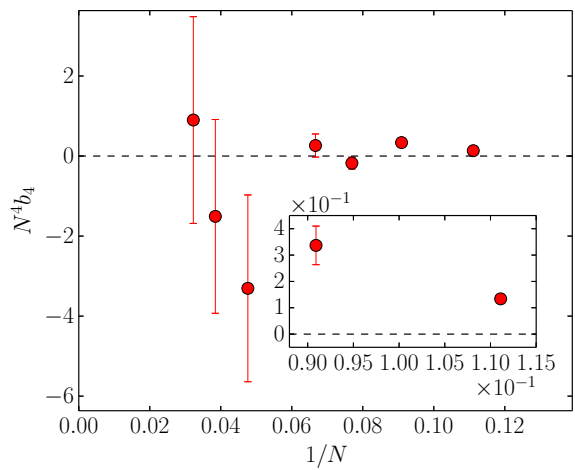}
\caption{Results obtained for {the} $b_4$ {coefficient in}
$CP^{N-1}$ models. }
\label{fig_grande_N_b_4}
\end{figure}

In the large $N$ limit, analytical computations~\cite{cpn_large_N,
SU(N)_large_N_limit} give for $b_4$ the prediction reported in
Eq.~\eqref{large_N}, hence we expect $b_4$ to be negative (and very small) for
large enough $N$. The $N$-dependence that is numerically observed presents
however some peculiar trends (Fig. \ref{fig_grande_N_b_4}).

\begin{figure}[htb!]
\includegraphics[scale=0.45]{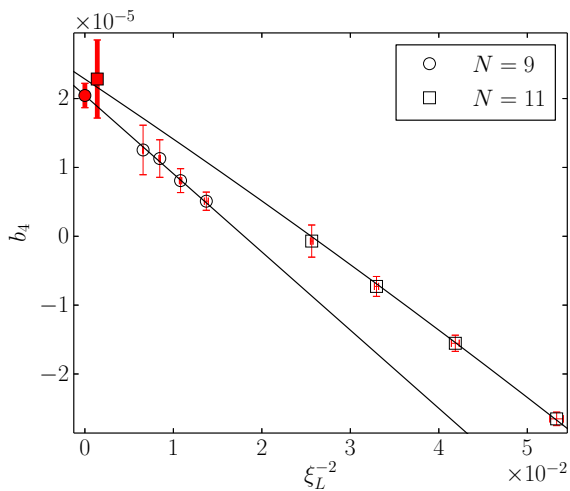}
\caption{Extrapolation to the continuum limit of $b_4$ for $N=9$ and $11$.
{Continuum values are reported as filled symbols and are slightly shifted
horizontally to increase readability.}}
\label{fig_continuum_limit_b_4_N_9_and_11}
\end{figure}

For $N\ge21$ negative values of $b_4$ are observed at finite lattice spacing, but
the continuum extrapolations are well consistent with zero. This is more or
less the behavior one could have guessed \emph{a priori} given the large $N$
theoretical results. However, one should notice that 
the $1.5~\sigma$ hint of a negative continuum result for
$N=21$ ($b_4=-1.7(1.2) \cdot 10^{-5}$) is in any case lower by more 
than one order of magnitude with respect to the leading large $N$ prediction
($b_4 \simeq - 7 \cdot 10^{-4}$ for $N = 21$).
Furthermore, this hint for a negative (continuum) $b_4$ does not
become a full evidence for smaller values of $N$ and in fact something new
happens for $N$ close to 10: for $N=9$ and $11$ the continuum extrapolations are
clearly positive, although at coarse lattice spacing negative $b_4$ values are
observed (see Fig. \ref{fig_continuum_limit_b_4_N_9_and_11}).

This fact suggests that, when looking at $b_4$, the large $N$ limit is not even qualitatively correct for small $N$
values.  An intriguing possibility, already advocated above, is that
the physics at small $N$ is in fact dominated by the small instantons
responsible for the singularities that are present in the 
$N=2$ case. It was shown
in the previous section that, assuming a weakly interacting ensemble of
instantons, this possibility leads to a peculiar small-$N$ behavior for
$\chi$, which is in agreement with numerical data. Using this approximation it
is simple to show that $f(\theta)$ has to be of the functional form
$f(\theta)=\chi(1-\cos\theta)$, just like in DIGA, from which one gets $b_2=-1/12$ and $b_4=1/360$.
While the numerical values obtained for $N=9$ and $11$ are still far from these
values, the qualitative picture we get is correct: $b_2$ is always negative (at
large $N$ due to Eq.~\eqref{large_N}, at small $N$ because $b_2=-1/12$ in the
DIGA) while $b_4$ changes sign. Also the behavior as a function of the lattice
spacing of $b_4$ is easily explained: for small $N$ it becomes positive only
when the lattice spacing is small enough for small instantons to become
dominant.

It is interesting to compare $CP^{N-1}$ models with 
$SU(N_c)$ Yang-Mills theories,
where the situation is different in several respects: on one hand, we do not
have analytical control on the large-$N_c$ limit, on the other hand nothing
singular happens in the small-$N_c$ case, and in particular for $N_c=2$. The
absence of small-$N_c$ singularities could be the reason, together with the fact
that the expansion is in powers of $1/N_c^2$, why the large $N_c$ 
scaling is known
to extend in this case down to very small $N_c$-values, i.e.~$N_c=3$ 
and even $2$.
Since previous studies performed in $SU(N_c)$ 
with $N_c \ge 3$ could not identify a
non-vanishing $b_4$ value, we here perform a dedicated study for $SU(2)$.

\begin{figure}[!htb]
\centering
\includegraphics[scale=0.45]{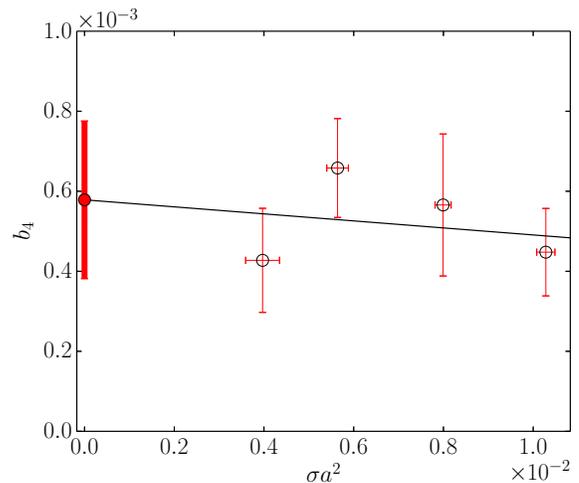}
\caption{Results obtained for $b_4$ in $SU(2)$ Yang-Mills theory, together with
their continuum extrapolation (reported as filled {symbol}).}
\label{fig_SU(2)_b_4}
\end{figure}

In order to estimate $b_4$ in $SU(2)$ Yang-Mills theory we performed
simulations using the same techniques described in Ref.~\cite{SU(N)_large_N_limit},
to which we refer for further details. Standard heath-bath \cite{Creutz:1980zw,
Kennedy:1985nu} and over-relaxation \cite{Creutz:1987xi} algorithms were adopted
to investigate four different values of the bare coupling ($\beta=2.70, \, 2.743, \,
2.7979, \, 2.85$); the corresponding lattice spacings ($a\sqrt{\sigma}\simeq
0.1014, \, 0.0894, \, 0.0751, \, 0.063$) were extracted from
Ref.~\cite{Fingberg:1992ju} or cubic interpolation of data thereof. For each value
of $\beta$ simulations were performed for seven values of $\theta_L$ in the
range $0\le \theta_L\le 12$ and the physical size of the lattice satisfied in
all the cases the relation $L\sqrt{\sigma}\gtrsim 3$. The numerical results
obtained using this set-up are shown in Fig.~\ref{fig_SU(2)_b_4}, from which we
see that lattice artefacts are smaller than our statistical errors and
extrapolating to the continuum limit with a linear function in $a^2$ we obtain
$b_4=6(2) \cdot 10^{-4}$.

{Despite the fact that the large-$N_c$ expansion works well also for small $N_c$
values, it would be far too optimistic to extract a large $N_c$ behavior from
just an $SU(2)$ result. It is nevertheless interesting to note that, 
assuming 
that the $1/N_c^4$ scaling holds down to $N_c  = 2$, one would expect 
$b_4 \sim 10^{-4}$ for $N_c = 3$, which is well within the upper
bound on $b_4$ which was possible to achieve  
in Refs.~\cite{SU(N)_large_N_limit} and \cite{theta_dep_SU3_d_elia}:
that clarifies 
why a non-vanishing value of $b_4$ was never detected in previous studies
performed for $N_c > 2$.}  

{The value that we have obtained is also 
quite reasonable when compared with other results in the
literature. 
For instance, when considering predictions about $\theta$-dependence
in pure gauge theories obtained within holographic 
approaches~\cite{Bigazzi:2015bna,Dubovsky:2011tu},
one obtains $b_4(\mathrm{HolYM})\simeq 0.033/N_c^4$, 
i.e.~$b_4 \simeq 2\cdot 10^{-3}$ in the two color 
case~\cite{Bigazzi:2015bna},
which has the same sign and is just a factor three larger than
our result. The same order of magnitude, even if with an opposite
sign, is obtained
using the expression for the vacuum energy $\theta$-dependence in QCD derived 
from chiral perturbation theory (see, e.g.,
Ref.~\cite{diCortona:2015ldu}): it is easy to see that in 
this case one gets
$b_4(\mathrm{QCD})=-3.5 \cdot 10^{-4}$.}

\section{Conclusions}
\label{section_conclusions}

The purpose of this study was to make progress towards
a numerical validation of analytic large-$N$ predictions for 
the $\theta$-dependence of 2$d$ $CP^{N-1}$ models.
To that aim, we have explored a range of $N$ going from
9 to 31, where we have provided continuum extrapolated results
both for $\xi^2 \chi$ and for the $b_2$ coefficient, which parameterize
fourth order terms in the $\theta$ expansion of the free energy
density. We have also been able
to provide, for the first time in the literature, a
determination of sixth order contributions, parameterized in terms of the 
$b_4$ coefficient, both for $CP^{N-1}$ models, with
$N \leq 11$, and for $SU(N_c)$ Yang-Mills theories with
$N_c = 2$.

The large-$N$ behavior of $\xi^2 \chi$ had already been explored 
by previous studies, which found deviations from the 
leading $1/N$ term of opposite sign with respect to 
the analytic prediction for the $1/N^2$ term~\cite{calcolo_e_2}. 
Our results have provided evidence that this may be due to 
a large $1/N^3$ contribution: indeed, once that is taken into account, 
numerical results are consistent with the prediction
of Ref.~\cite{calcolo_e_2}. 
A similar scenario applies to $b_2$, which has been investigated
systematically in our study for the first time: 
while, on one hand, the fact that $b_2$ is proportional 
to $1/N^2$ at the
leading order is well supported by numerical data, on the other
hand consistency also with 
the predicted prefactor appearing in Eq.~(\ref{b2largen}) 
is possible only when $O(1/N^4)$ corrections are taken into account.

The fact that higher order corrections in the $1/N$ expansion
seem to be significant for $N \sim O(10)$ is at odds with what is
observed for $SU(N_c)$ Yang-Mills theories. We have put that 
in connection with the expected finite radius of convergence of the series,
which is bounded by $1/2$ because of the divergence 
of the topological susceptibility taking place for $N = 2$,
where the topological fluctuations are dominated by small
instanton and antiistantons, whose density is UV divergent.

Having that in mind, we have tried to improve the convergence 
of the $1/N$ expansion, after guessing the leading
divergent behavior around $N = 2$. Indeed,
assuming such small topological objects to be weakly interacting,
one obtains a prediction for a $1/(N-2)$ divergence of the topological
susceptibility, which we have verified to be well supported 
by our present numerical data for $N < 15$.
We have therefore tried to expand the function 
$(N - 2) \xi^2 \chi$ as a regular series in $1/N$,
obtaining results which are marginally better and more stable 
than those obtained within the standard $1/N$ expansion for 
$\xi^2 \chi$.

The values obtained for the $b_4$ coefficient at $N = 9$ and 11,
which are of opposite sign with respect to the leading 
$1/N^4$ prediction,  are also consistent with the presence of significant
corrections related to small-$N$ physics.

Present results could be improved in the future in various directions. 
On one hand, an effort to extend the analysis to larger values 
of $N$, using for instance the algorithm proposed in 
Ref.~\cite{fighting_slowing_down_cpn}, could be useful especially
for a better quantitative test of the coefficient in front of the 
leading $1/N^2$ term for $b_2$; we have estimated that
$N \gtrsim 50$ should be explored to that purpose.
On the other hand, a precise determination of 
$\theta$-dependence for small values of $N$, approaching $N = 2$,
could be useful for a more precise matching between 
small-$N$ and large-$N$ behaviors. {It would be also interesting
to consider the non-zero momentum components of the 
topological susceptibility~\cite{Abe:2018loi}, which could 
give more information on the actual topological charge
distribution.}

Finally, let us comment on the result obtained 
for the $b_4$ coefficient in the 4$d$ $SU(2)$ pure gauge theory,
$b_4=6(2) \cdot 10^{-4}$. Assuming that the $1/N_c^4$ scaling holds
down to $N_c = 2$, this value is well within previous upper bounds
set for $b_4$ in $SU(N_c)$ gauge theories with $N_c \geq 3$, and
also well explains why a clear non-zero signal
has not yet been achieved for this quantity in those cases. 
At the same time, the value appears in good semi-quantitative 
agreement (i.e.~within a factor 3) with large-$N_c$ predictions
obtained within holographic models~\cite{Bigazzi:2015bna}.

\acknowledgments
It is a pleasure to thank F.~Bigazzi, G.~Dunne, 
K.~Fukushima, A.~Jevicki, P.~Rossi, 
F.~Sanfilippo and E.~Vicari for
useful discussions and comments.  Numerical simulations have been performed at
the Scientific Computing Center at INFN-PISA and on the MARCONI machine
at CINECA, based on the agreement between INFN
and CINECA (under project INF18\_npqcd).


\end{document}